\documentclass[sigconf]{acmart}
\usepackage{amsmath}
\usepackage{graphicx} 
\usepackage{caption} 
\usepackage{subfigure}
\usepackage{multirow}
\usepackage{algorithmic}
\usepackage{algorithm}
\usepackage{lipsum} 
\usepackage{fancyhdr} 
\usepackage{balance}

\copyrightyear{2025}
\acmYear{2025}
\setcopyright{acmlicensed}
\acmConference[KDD '25] {Proceedings of the 31st ACM SIGKDD Conference on Knowledge Discovery and Data Mining V.2}{August 3--7, 2025}{Toronto, ON, Canada.}
\acmBooktitle{Proceedings of the 31st ACM SIGKDD Conference on Knowledge Discovery and Data Mining V.2 (KDD '25), August 3--7, 2025, Toronto, ON, Canada}
\acmISBN{979-8-4007-1454-2/25/08}
\acmDOI{10.1145/3711896.3737156}

\begin{document}

\title{Time Matters: Enhancing Sequential Recommendations with Time-Guided Graph Neural ODEs}

\author{Haoyan Fu}
\authornote{Also with Guangdong Laboratory of Artificial Intelligence and Digital Economy (SZ).}
\email{haoyan-fu@bit.edu.cn}
\affiliation{%
  \institution{Beijing Institute of Technology}
  \city{Beijing}
  \country{China}
}

\author{Zhida Qin}
\authornote{Corresponding authors.}
\email{zanderqin@bit.edu.cn}
\affiliation{%
  \institution{Beijing Institute of Technology}
  \city{Beijing}
  \country{China}
}

\author{Shixiao Yang}
\email{ysx144_51@bit.edu.cn}
\affiliation{%
  \institution{Beijing Institute of Technology}
  \city{Beijing}
  \country{China}
}

\author{Haoyao Zhang}
\email{zhanghaoyao@bit.edu.cn}
\affiliation{%
  \institution{Beijing Institute of Technology}
  \city{Beijing}
  \country{China}
}

\author{Bin Lu}
\email{robinlu1209@sjtu.edu.cn}
\affiliation{%
  \institution{Shanghai Jiao Tong University}
  \city{Shanghai}
  \country{China}
}

\author{Shuang Li}
\email{shuangliai@buaa.edu.cn}
\affiliation{%
  \institution{Beihang University}
  \city{Beijing}
  \country{China}
}

\author{Tianyu Huang}
\email{huangtianyu@bit.edu.cn}
\affiliation{%
  \institution{Beijing Institute of Technology}
  \city{Beijing}
  \country{China}
}

\author{John C.S. Lui}
\email{cslui@cse.cuhk.edu.hk}
\affiliation{%
  \institution{The Chinese University of Hong Kong}
  \city{Hong Kong}
  \country{China}
}

\renewcommand{\shortauthors}{Haoyan Fu et al.}

\begin{abstract}
Sequential recommendation (SR) is widely deployed in e-commerce platforms, streaming services, etc., revealing significant potential to enhance user experience. The core of SR lies in exploring the sequential relationships in historical user-item interactions. However, existing methods often overlook two critical factors: \textit{irregular user interests} between interactions and \textit{highly uneven item distributions} over time. The former factor implies that actual user preferences are not always continuous, and long-term historical interactions may not be relevant to current purchasing behavior. Therefore, relying only on these historical interactions for recommendations may result in a lack of user interest at the target time. The latter factor, characterized by peaks and valleys in interaction frequency, may result from seasonal trends, special events, or promotions. These externally driven distributions may not align with individual user interests, leading to inaccurate recommendations. To address these deficiencies, we propose TGODE to both enhance and capture the long-term historical interactions. Specifically, we first construct the user time graph and item evolution graph, which utilize user personalized preferences and global item distribution information, respectively. To tackle the temporal sparsity caused by irregular user interactions, we design a time-guided diffusion generator to automatically obtain an augmented time-aware user graph. Additionally, we devise a user interest truncation factor to efficiently identify sparse time intervals and achieve balanced preference inference. After that, the augmented user graph and item graph are fed into a generalized graph neural ordinary differential equation (ODE) to align with the evolution of user preferences and item distributions. This allows two patterns of information evolution to be matched over time. Experimental results demonstrate that TGODE outperforms baseline methods across five datasets, with improvements ranging from 10\% to 46\%. The code is available
at https://github.com/Qin-lab-code/TGODE.

\end{abstract}

\ccsdesc[500]{Information systems~Recommender systems}

\keywords{Sequential Recommendation; Graph Neural Networks; Neural Ordinary Differential Equation; Diffusion Model}

\maketitle

\section{Introduction}
Recommender systems, as the method of delivering personalized preference recommendations to users, effectively address the issue of information overload in the modern Internet era. Furthermore, many research efforts \cite{ma2020memory, li2020time, fan2022sequential} have recognized that user preference information is partially embedded within dynamic sequential item access behaviors, which exhibit strong correlations with preference information. Consequently, the task of sequence recommendation (SR) can efficiently utilize these correlations to explicitly model user sequential behavior.

Numerous approaches have been proposed for SR tasks to capture users' sequential interaction histories. Early efforts employ traditional methods, such as Markov chains \cite{rendle2010factorizing} and RNNs \cite{hidasi2015session,quadrana2017personalizing}, to explore long-term sequence dependencies. With the success of transformers, a new wave of self-attentive SR models \cite{hou2022core,zhang2023adaptive,ye2023graph} is developed. However, these models often face limitations due to the inherent data sparsity in recommendation systems. Recently, Stochastic Differential Equation (SDE)-based probabilistic models have gained traction as a means to enhance existing SR methods. Diffusion models, such as DiffRec \cite{wang2023diffusion} and DreamRec \cite{yang2024generate}, have emerged as promising tools for modeling complex interaction generation. Additionally, neural ODE \cite{guo2022evolutionary,qin2024learning} has been applied within the recommendation domain to model continuous sequential interactions, further advancing the capabilities of SR systems.

\begin{figure}[t]
	\centering
	
		\centering
		\includegraphics[width=1.0\linewidth]{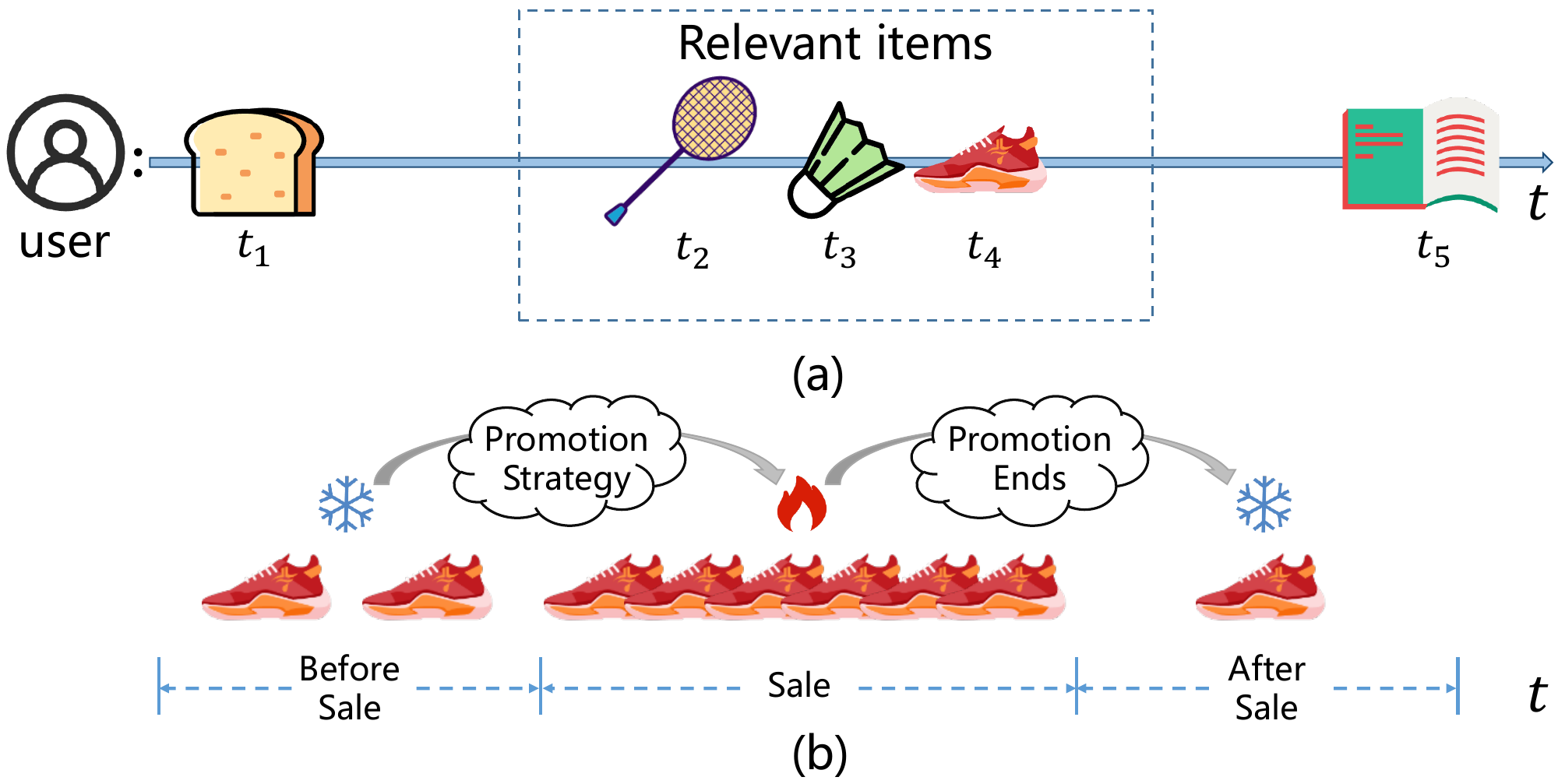}
	
        \caption{Illustration example of sequential recommendation with (a) irregular user interests and (b) highly uneven item distributions.}
    
 \label{fig:intro}
\end{figure}

Existing sequential recommendation methods, while effective in capturing sequential relationships in user sequences, ignore two critical factors: (i) \textbf{Irregular user interests.} General sequential modeling assumes regular user interactions over time, which is inconsistent with real-world recommender systems. As shown in Figure 1, user interactions follow an irregular distribution. Users may purchase closely related items consecutively in short time intervals when the sequential relationship of items represents the current user preferences. Comparatively, items outside of long time intervals are irrelevant to the current purchase behavior. Therefore, capturing irrelevant sequential patterns in a straightforward manner can have a negative effect. (ii) \textbf{Highly uneven item distributions.} Referring to Figure 1, the popularity of an item can change over time due to external factors. For example, the sales of an item may be highly increased due to a promotional campaign, and this effect is independent of the user's personal preferences. That is, the user's current behavior may be influenced by external promotions rather than following personal preferences. However, such outside influences over time cannot be mined by historical interactions, which limits the ability to make accurate recommendations. 

Despite their great importance, the above two phenomena indicate that user interests and item distributions are both highly time-dependent with irregular intervals. Hence, incorporating this information into the SR system is a non-trivial problem. There are two challenges to deal effectively with both of these types of information. The first challenge is to alleviate the time-imbalanced user interactions. User behaviors tend to be highly concentrated in a short period of time and are absent in most timestamps, which exhibits the temporal sparsity issue. Most existing works mainly utilize the augmentation methods (such as random dropouts, similar sequences clustering) to supplement the interactions between cold start users and items from those similar ones with enough interactions. However, due to heterogeneous dynamics of user preference on time dimension, simply copying those interactions from similar ones and ignoring the time-aware interests may lead to severe user preference deviation and reduced performance. The second challenge lies in the mismatch between the distribution of personalized user interests and target items over time. As we demonstrate in the data analysis (See Section 2.2), the item distribution over time is mainly driven by some external factors, i.e., the promotions or advertising, which is independent of the evolution of user preferences. Failing to consider such phenomenon will make users crowded by disliked items and inevitably result in suboptimal recommendation results. Hence, it is necessary to characterize the evolution processes of both user individual preference and item distributions to efficiently align them and improve recommendation accuracy.

To address the above challenges, we propose a novel sequential recommendation framework with a \textbf{T}ime-\textbf{G}uided graph neural \textbf{ODE}s, named \textbf{TGODE}. First, we construct two tailored graphs, i.e. user time graph and item evolution graph, to capture individual user preference changes and explore dynamic item distributions respectively. Second, we designed a time-guided diffusion generator to augment the user time graph. Specifically, we apply a VAE encoder to compress interactions into latent vectors, and the output of another ODE module is co-coded to provide additional sequential information. We then perform a novel temporal embedding encoder to guide the diffusion process. Guided by temporal embedding, the diffusion model is able to recover user interactions corresponding to the time. Further, we propose a user preference inference module to carry out the generation process. Accordingly, we can generate potential user interests and achieve temporally balanced user preferences. Then, we propose a generalized graph neural ODE for matching critical evolutionary information in both the augmented user graph and original item graph. Specifically, our graph neural ODE could align the evolutionary patterns of two graphs over time. In this way, user interest transitions are matched with item trends on the timeline. In our framework, the outputs of these two modules serve as mutual inputs, and an iterative training strategy is employed to enhance their performance. We conduct extensive experiments on five real-world datasets and validate that our TGODE outperforms multiple state-of-the-art baselines.

Our contributions can be summarized as follows:
\vspace{-5pt}
\begin{itemize}
\item We propose TGODE, a novel sequential recommendation framework that captures irregular user interests and highly uneven item distributions using two tailored graphs: a user time graph and an item evolution graph.

\item We develop a time-guided diffusion generator to augment the user time graph, enabling the model to recover and generate user interactions that are temporally balanced and reflective of potential user preferences.

\item Our generalized graph neural ODE aligns the temporal evolution of user interests with item trends, ensuring consistent and accurate recommendations over time.

\item Extensive experiments on five real-world datasets show that TGODE outperforms state-of-the-art baselines, confirming its effectiveness in sequential recommendation tasks.
\end{itemize}

\section{PRELIMINARY}

\subsection{Problem Definition}

The traditional approach to sequence recommendation involves predicting the next item in a user interaction sequence $s$, denoted as $s = \{ (v_1, t_1), (v_2, t_2), \ldots, (v_n, t_n) \}$, where $n$ represents the length of sequence $s$, and $(v, t) \in s$ signifies an interaction between the sequence and item $v$ at time $t$. Typically, each item $v$ is initiated as its corresponding embedding $\mathbf{x}$. The objective is to forecast the $(n+1)$-th potential interaction given the first $n$ interacting items and the target time $t_{n+1}$.

\subsection{Data Analysis}
In order to thoroughly investigate the widespread issues of irregular user interests and highly uneven item distributions in real-world scenarios, we conduct an extensive data analysis. For this study, we use two datasets from Amazon: Beauty and Toys. These are subsets of the extensive Amazon Review Data, and detailed statistics for both datasets are provided in the Appendix \ref{app:dataset}. For simplicity, we present the data analysis results for the beauty dataset in the main text, while the results for the toys dataset are provided in the Appendix \ref{app:DataAnalysis}.

\subsubsection{Analysis of Irregular User Interests}
A time interval represents the arbitrary time difference between two consecutive interactions for a user, reflecting the temporal connection between these interactions. We quantify the proportions of all pre-existing time intervals across different ranges in both datasets, as shown in Figure \ref{a}. Our observations reveal that, although nearly half of the interactions occur instantaneously, there were still numerous delayed interactions. The proportion of interactions across various time intervals is significant, indicating substantial temporal irregularity for the user. Almost 15\% of the delayed interactions occur after more than 350 time intervals. These extremely long time intervals suggest that such interactions may only be weakly related to previous interactions. However, previous sequential modeling techniques often fail to distinguish these delayed interactions from those occurring instantaneously, potentially leading to biased interpretations of user preferences.

To further explore the relationship between interactions and time intervals for individual users, we randomly select three representative users and visualize their interaction timelines in different colors, as shown in Figure \ref{b}. The figure demonstrates that these users' interactions are clustered within certain time ranges and irregularly distributed at other timestamps. This irregular distribution underscores the inadequacy of purely sequential models in capturing time interval-related personalized interests, highlighting the necessity of considering irregular time intervals in modeling.

\subsubsection{Analysis of Highly Uneven Item Distributions}
\begin{figure}[t]
	\centering
	\subfigure[Proportions of Time Intervals ]{
		\centering
		\includegraphics[width=0.45\linewidth]{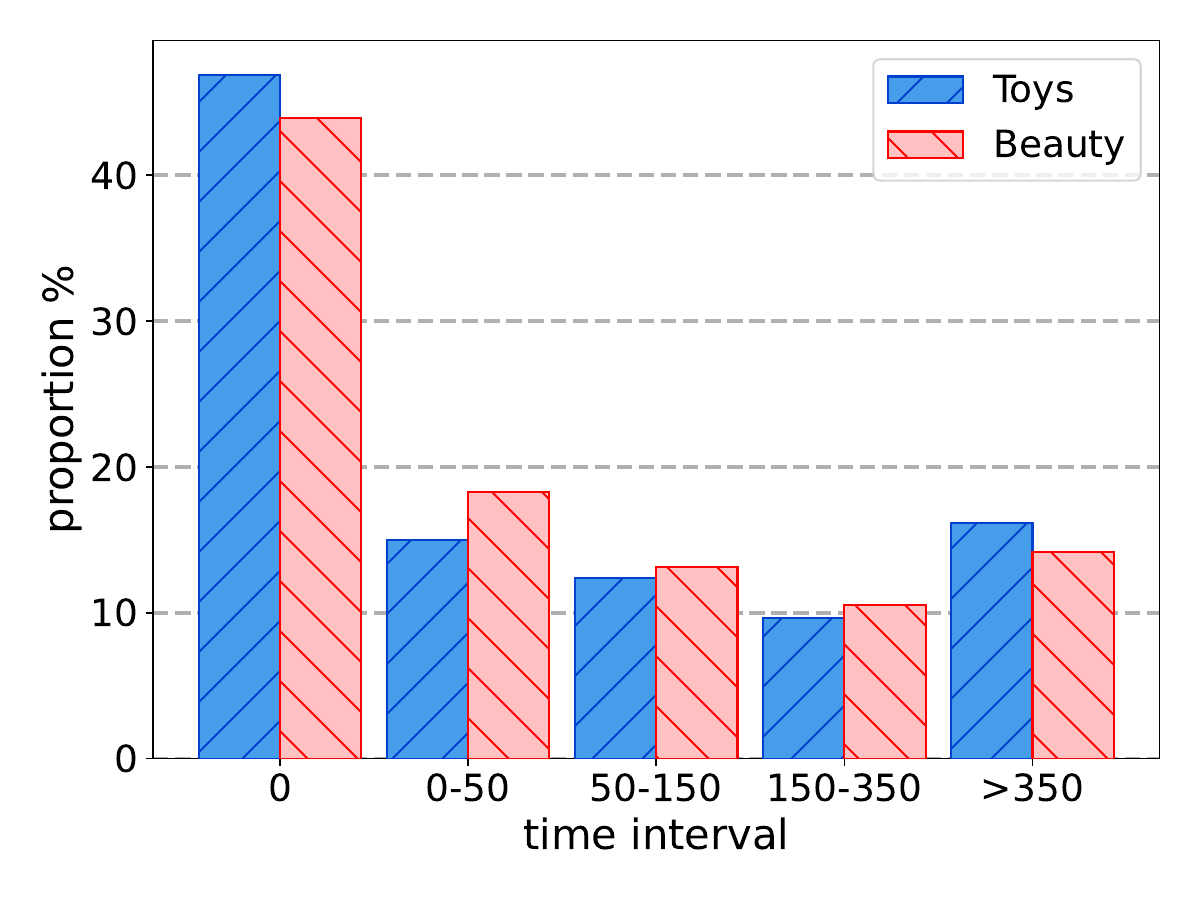}
		\label{a}
	}%
        \subfigure[User Interaction Timelines ]{
		\centering
		\includegraphics[width=0.45\linewidth]{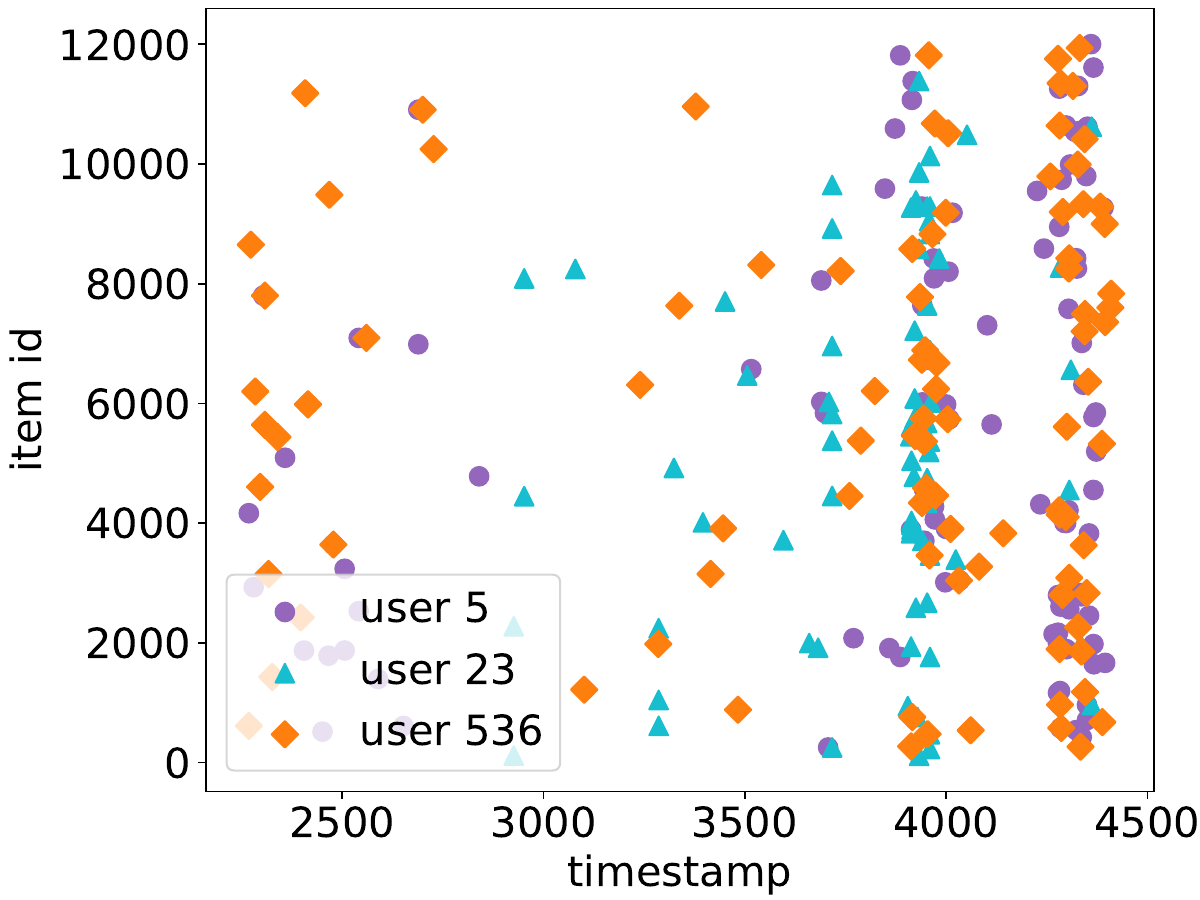}
		\label{b}
	}\\
    \vspace{-0.3cm}
        \subfigure[Item Emergence Ratios ]{
		\centering
		\includegraphics[width=0.45\linewidth]{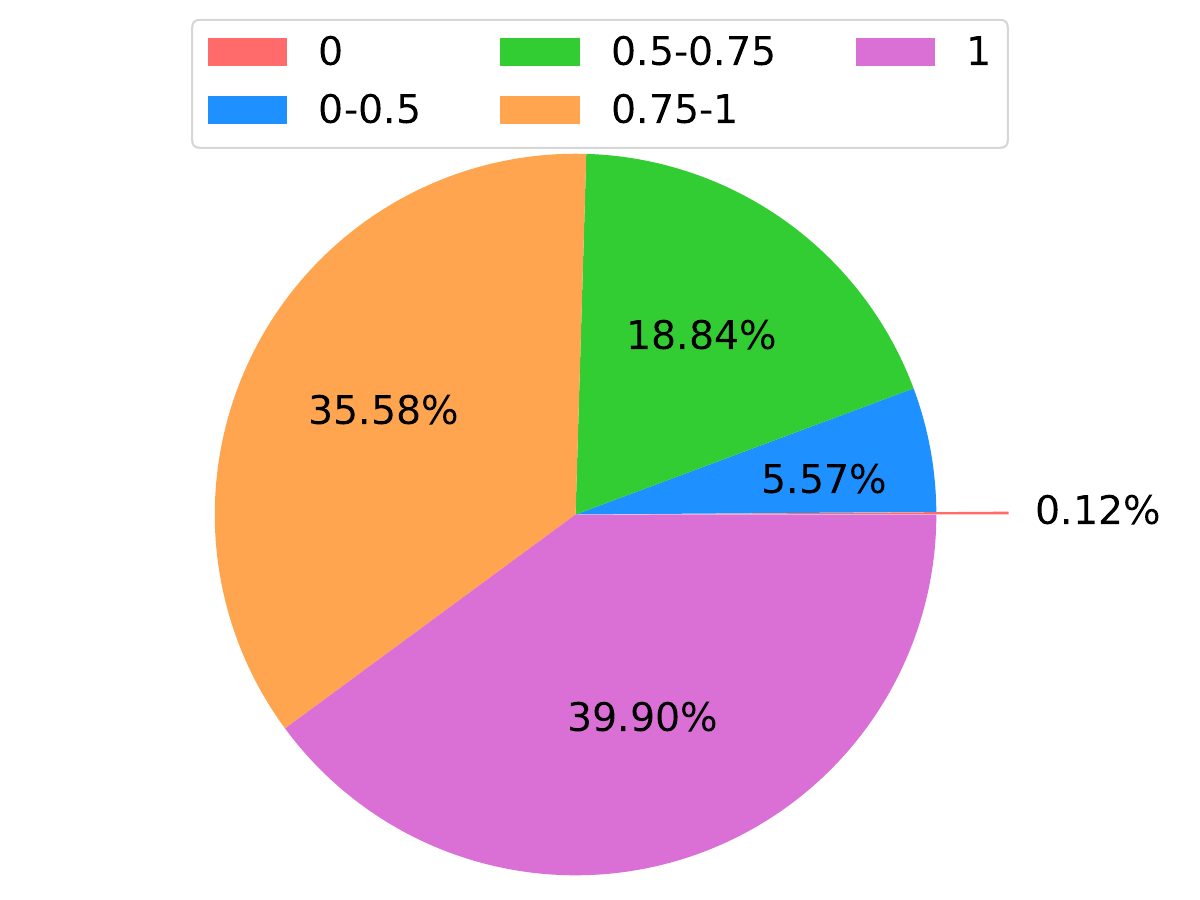}
		\label{c}
	}%
        \subfigure[Item Interaction Timelines ]{
		\centering
		\includegraphics[width=0.45\linewidth]{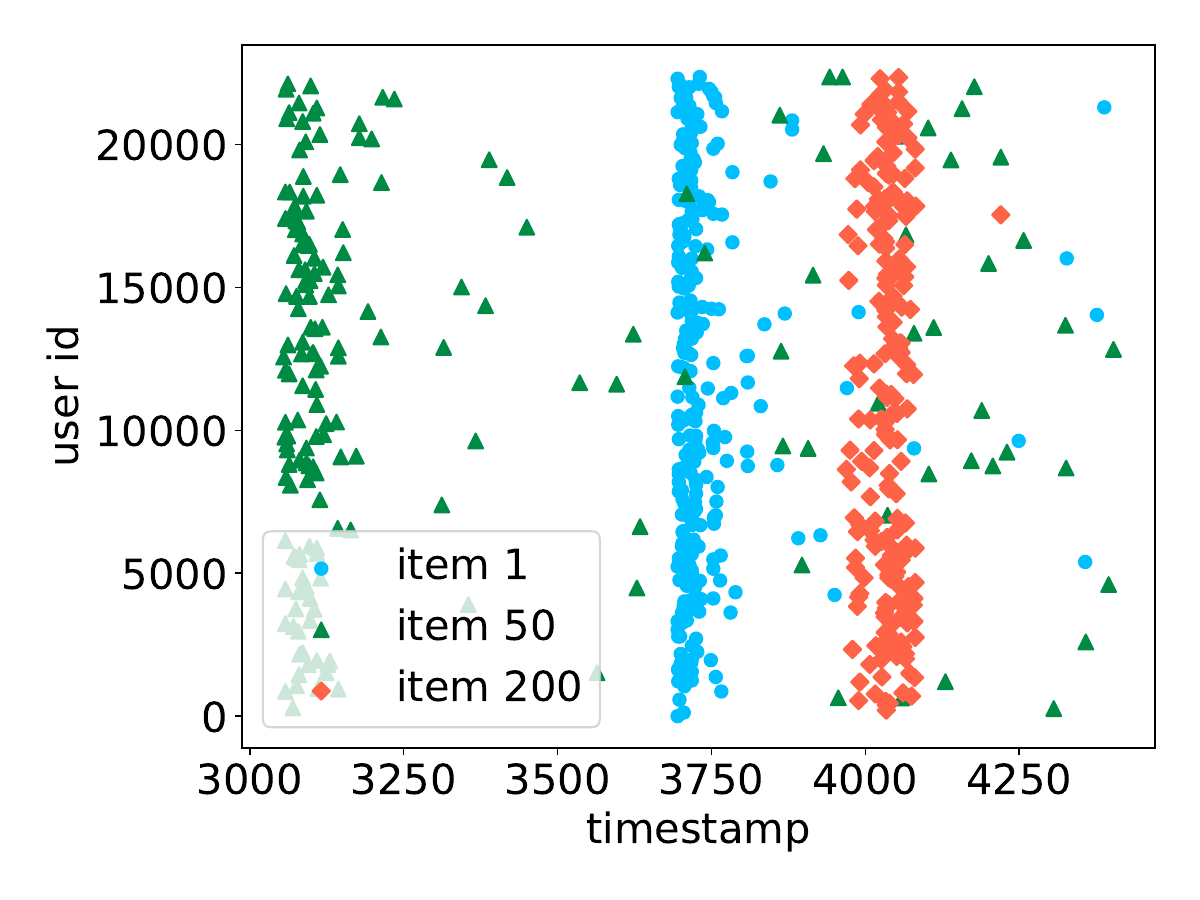}
		\label{d}
	}%
    \vspace{-0.3cm}
 
        \caption{Data analysis regarding the Beauty dataset.}

 \label{fig:datas1}
\end{figure}

In real-world scenarios, user interests evolve over time, and items themselves are subject to changes in distributions due to exposure, promotions, and other factors. To quantify this, we analyze the changes in item interactions over time. We divide the timeline into 250-day length slices and calculate the proportion of an item's interactions within each slice to assess the item's concentrated emergence over time. In fact, the higher the proportion of emergence, the more concentrated the distribution of the item on the timeline is around a specific time, that is, the more popular the item is during a certain period. We categorize these proportions into the ranges of \{0, 0-0.5, 0.5-0.75, 0.75-1, 1\}. As illustrated in Figure \ref{c}, over 75\% of items exhibit an emergence ratio of more than 75\%, while items with an emergence ratio of less than 50\% account for less than 10\% of all items. This analysis demonstrates that items are highly concentrated within time slices, which we refer to as highly uneven item distributions.

To visualize item popularity more effectively, we select three items and display their interaction timelines, as shown in Figure \ref{d}. These visualizations confirm that items interact with numerous users around specific timestamps, corroborating the findings in Figure \ref{c}. However, this prevalent highly uneven item distribution is often overlooked and cannot be adequately captured by traditional time-series methods. Therefore, it is crucial to model data in a time-aware manner to accurately reflect these dynamics.

\begin{figure*}[htbp]
    \centering
    \includegraphics[width=0.85\linewidth]{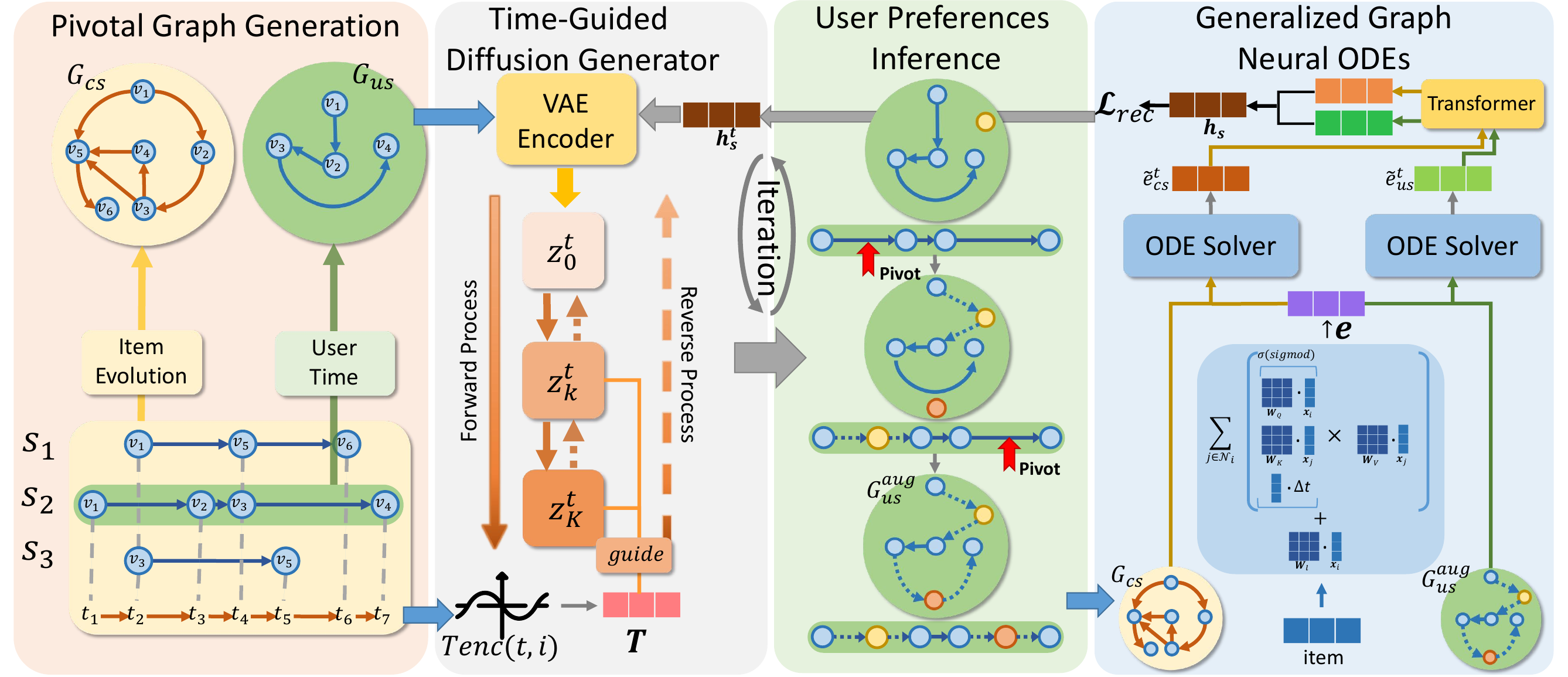}
    \caption{The overall architecture of our proposed method.}
    \label{fig:main}
\end{figure*}

\section{Method}

\subsection{Continuous Time Evolution Process}

To capture the nuanced dynamics of user preferences over time, it is imperative to model continuous item interactions effectively. To this end, we devise a continuous time process, illustrated in Figure \ref{fig:main}. Beginning with a collection of all user sequences, $\mathbf{\mathcal{S}} = \{s_1, s_2, \ldots, s_{\left| \mathbf{\mathcal{S}} \right|}\}$, we integrate these sequences along a unified timeline, aligning their temporal occurrences.

However, it is clear from the data analysis that the unbalanced time intervals appear on both the individual user sequences and the overall item distribution, so it is necessary to construct the two graph structures separately to mine the corresponding information. Concretely, we construct two pivotal graphical structures:
\begin{itemize}
    \item User Time Graph ($\mathbf{\mathcal{G}}_{us}$). This graph captures individual user preferences. It is defined as $\mathbf{\mathcal{G}}_{us} = \{\mathbf{\mathcal{V}}_{us}, \mathbf{\mathcal{E}}_{us} \}$. Here, $\mathbf{\mathcal{V}}_{us} = \{v_i | v_i \in s\}$ represents all nodes corresponding to items within each user sequence $s$, while $\mathbf{\mathcal{E}}_{us} = \{(v_i, v_j, t) | v_i, v_j \in s\}$ denotes the set of transitional relationships between two items within a sequence, labeled with their corresponding interaction times.
    \item Item Evolution Graph ($\mathbf{\mathcal{G}}_{cs}$). This graph delineates the overarching temporal representation of items across all sequences. It is structured as $\mathbf{\mathcal{G}}_{cs} = \{\mathbf{\mathcal{V}}_{cs}, \mathbf{\mathcal{E}}_{cs} \}$. In this case, $\mathbf{\mathcal{V}}_{cs} = \mathbf{\mathcal{I}}$ includes the entirety of items, while $\mathbf{\mathcal{E}}_{cs} = \{(v_i, v_j, t) | v_i, v_j \in \mathbf{\mathcal{S}}\}$ represents interactions occurring at time $t$ across the entire sequence collection $\mathbf{\mathcal{S}}$.
\end{itemize}
The distinctive focus of these graphs lies in their respective emphasis on individual user preferences and global item distributions. In particular, the graph $\mathbf{\mathcal{G}}_{us}$ that models user preferences is more representative of current user interests. However, the number of interactions of the user graph $\mathbf{\mathcal{G}}_{us}$, which only considers the sequence of individual behaviors, is much smaller than that of the evolution graph $\mathbf{\mathcal{G}}_{cs}$. This leads to severe sparsity in the time dimension. Therefore, it is necessary to design a time-guided user preferences generator to fill in the missing potential user interactions. In this generative pattern, the potential user interest transition process can be obtained between long intervals of interactions.

\subsection{Time-Guided Graph Neural ODEs}

To efficiently extract temporal information from the two constructed graphs, we design a time-guided graph neural ODE model. The model consists of three parts: 1) a time-guided diffusion generator to augment user time graphs; 2) a user preferences inference to generate potential user interests; and 3) a generalized graph neural ODE to match evolution information on both graphs.

\subsubsection{\textbf{Time-Guided Diffusion Generator}}
\ 
\newline
Due to irregular user interests, user time graphs are often sparse. Therefore, a generator is needed to address this sparsity in the temporal dimension. Inspired by the ability of diffusion models to retain detailed information during generation, we propose a novel time-specific approach. Specifically, we design a time-variant diffusion module to generate user behavior interactions over extended time intervals. Our approach focuses on creating potential transition paths for interacting items, thereby enhancing continuous user interest preferences.

To achieve this, we employ a time-guided denoising probabilistic model to learn time-specific user interaction information. We get the potential presentations with a vector encoder. Then we progressively disrupt the potential vectors and use the encoded temporal embedding to recover the original interaction records corresponding to the time. After training the model, a user preference inference strategy is utilized to generate an augmented user graph. This strategy effectively addresses the sparsity of user interactions over time and appropriately augments the missing edges in the original graph.

\noindent \textbf{The Latent Vectors Encoder.} First, we need to model the initial inputs $\mathbf{z}^t_0$ in order to perform the forward and backward processes. However, the user time graph only contains the current user's interactions, which is not sufficient for reasoning about potential interactions at other times. Inspired by the latent features encoder in latent diffusion \cite{rombach2022high}, we use Variational Autoencoder (VAE) to compress graph structure and sequence information into the potential space at time $t$:
\begin{equation}
\label{0}
\begin{aligned}
& \mathbf{z}^t_0 = \text{VAE}(\mathbf{A}_{us}^t, \mathbf{h}_s^t),  
\end{aligned}
\end{equation}
where $\mathbf{A}_{us}^t$ is the adjacency matrix of the graph $\mathbf{\mathcal{G}}_{us}$ at time $t$ and $\mathbf{h}_s^t$ is the sequence representation modeled by the subsequent graph neural ODE process. It is worth noting that the sequence information should match the current time $t$ and be able to recover information from other moments. Our ODE process can model the representation of continuous time, thus helping the graph generation at other moments.

\noindent \textbf{The Temporal Embedding Encoder.} In the original user graph, each item node has a defined interaction time. To accurately capture the timing of user interactions, we transform timestamps into corresponding time embeddings. We design a temporal embedding encoder to derive the temporal guidance for the diffusion model. Specifically, we combine sine and cosine values of varying frequencies with nonlinear transformations to encode time as follows:
\begin{equation}
\label{1}
\begin{aligned}
& \mathbf{c}_t = \text{Concat}\left(\sin\left(2\pi\omega_i t + b_i\right), \cos\left(2\pi\omega_i t + b_i\right)\right),  
\end{aligned}
\end{equation}
where $\omega_i$ is the first $i$ frequency and $b_i$ is the offset term. 

\noindent \textbf{Time-Guided Diffusion Process.} Using the encoded temporal embedding, the diffusion model meticulously carries out the process of user graph corruption and reconstruction through forward and reverse processes. Specifically, we progressively disrupt the initial state $\mathbf{z}^t_0$ through $K$ steps, known as the forward process. Following DDPM \cite{ho2020denoising}, we parameterize the forward process from $\boldsymbol{z}_0^t$ to $\boldsymbol{z}_{K}^t$ as follows:
\begin{equation}
\label{2}
\begin{aligned}
    q(\boldsymbol{z}_{K}^t |\boldsymbol{z}_{0}^t) = \mathcal{N}(\boldsymbol{z}_{K}^t; \sqrt{\bar{\alpha}_{K}^t }\boldsymbol{z}^t_{{K}-1}, (1-\bar{\alpha}_{K})\boldsymbol{I});\\
    \boldsymbol{z}^t_{{K}}=\sqrt{\bar{\alpha}_{{K}}} \boldsymbol{z}_{0}^t+\sqrt{1-\bar{\alpha}_{{K}}} \boldsymbol{\epsilon}, \quad  \boldsymbol{\epsilon} \sim \mathcal{N}(\boldsymbol{0}, \boldsymbol{I}),
\end{aligned}
\end{equation}
where $\alpha_{{K}}=1-\beta_{{K}}, \bar{\alpha}_{{K}}=\prod_{k=1}^{{K}} \alpha_{k}$, for $1-\bar{\alpha}_{{K}} \propto {K}$.

After the forward process corrupts the original graph structure into Gaussian noise, we control the reverse process by the learned time embedding $\mathbf{c}_{t}$. This process starts with $\boldsymbol{z}_{K}^t$ and gradually removes noise through a neural network to restore the user graph:
\begin{equation}
\label{3}
    p_{\theta}(\boldsymbol{z}_{k-1}^t|\boldsymbol{z}_{k}^t, \mathbf{c}_{t}) = \mathcal{N}(\boldsymbol{z}_{k-1}^t; \boldsymbol{\mu}_{\theta}(\boldsymbol{z}_{k}^t,\mathbf{c}_{t}, k), \boldsymbol{\Sigma}_\theta{(\boldsymbol{z}_{k}^t,\mathbf{c}_{t},k)}),
\end{equation}
where the terms $\boldsymbol{\mu}_{\theta}(\boldsymbol{z}^t_{k},\mathbf{c}_{t},k)$ and $\boldsymbol{\Sigma}_\theta{(\boldsymbol{z}^t_{k},\mathbf{c}_{t},k)}$ represent Gaussian parameters generated by neural networks.

\noindent \textbf{Time Smoothness Optimization.} To optimize the parameter $\theta$ within the neural network, we maximize the Evidence Lower Bound (ELBO) associated with the initial state $\boldsymbol{z}_0^t$. After derivation of the formula, the loss function for step $k$ is as follows:
\begin{equation}\label{4}
\begin{aligned}
     \mathcal{L}_k & = 
      \mathbb{E}_{q(\boldsymbol{z}_k^t|\boldsymbol{z}_0^t)}\left[D_{KL}(q(\boldsymbol{z}^t_{k-1}|\boldsymbol{z}_k^t,\boldsymbol{z}_0^t))
\parallel p_\theta(\boldsymbol{z}_{k-1}^t|\boldsymbol{z}_k^t, \mathbf{c}))\right] \\
     & = \mathbb{E}_{q(\boldsymbol{z}_k^t|\boldsymbol{z}_0^t)}
\left[\frac{1}{2} \left(\frac{\bar{\alpha}_{k-1}}{1-\bar{\alpha}_{k-1}}-\frac{\bar{\alpha}_{k}}{1-\bar{\alpha}_{k}}  \right){\lVert {\hat{\boldsymbol{z}}_\theta(\boldsymbol{z}_k^t, \mathbf{c}, k)-\boldsymbol{z}_0^t} \rVert }_2^2\right] + C,
\end{aligned} 
\end{equation}
where $\hat{\boldsymbol{z}}_\theta(\boldsymbol{z}_k^t, \mathbf{c}, k)$ is obtained by feeding the input vector $\boldsymbol{z}_k^t$ and the time embedding of step $k$ into a MLP to predict $\boldsymbol{z}_0^t$.

Considering Eq. (\ref{4}), we can optimize the ELBO by utilizing $\sum^{K}_{k=1}\mathcal{L}_k$. In practical implementation, we uniformly sample step $k \sim \mathcal{U}(1, K)$ for optimizing $\mathcal{L}(\boldsymbol{z}_0^t, \theta)$, formalized as follows:
\begin{equation}\label{5}
    \mathcal{L}(\boldsymbol{z}_0^t, \theta) = \mathbb{E}_{k \sim \mathcal{U}(1, K)}\mathcal{L}_k.
\end{equation}

Although this optimization ensures that the restored $\hat{\boldsymbol{z}}_0^t$ is as close as possible to the original state $\boldsymbol{z}_0^t$, it may lead to over-smoothing of time. Since most of the original item interactions are within the same time period, the model tends to generate nodes within repeated timestamps. This makes it difficult for the model to learn user behavior under sparse timestamps, which is not conducive to generating high-quality samples. To mitigate this issue, we regularize the edge weights to increase the irregularity of the graph indirectly. The overall loss function of our time-guided diffusion module is:
\begin{equation}\label{6}
    \mathcal{L}_{diff} = \mathcal{L}(\boldsymbol{z}_0^t, \theta) + \sum_{i, j}\left|\hat{\boldsymbol{z}}_0^{t}\right|.
\end{equation}

\subsubsection{\textbf{User Preferences Inference}}
\ 
\newline
After training a time-guided diffusion model, a key question is how to infer the potential interest of users at other times. On the timeline, there are continuous interactions during some time periods, while others may be almost absent. We devise a criterion, called the user interest truncation factor, to identify time periods where user interest is missing.

\noindent \textbf{User Interest Truncation Factor.} Specifically, given an overall timeline $T$, we divide it into $m$ time pivots $t^p = \{t^p_1, t^p_2, \ldots, t^p_m\}$. Assuming the original time series $t^o = \{t^o_1, t^o_2, \ldots, t^o_n\} $, for each timestamp $t^o_i$, we define the function $g(t^o_i)$ to find the closest time pivot $t^p_j$:
\begin{equation} \label{7}
    g(t^o_i) = \mathop{\arg\min}\limits_{t^p_j \in t^p} \left| t^p_j - t^o_i \right|.
\end{equation}
Based on the above function, we can get the set of uncovered pivots $t^p_{set} = \{t \in t^p \mid t \notin g(t^o_i)\}$. This set contains the time pivots where the user's interest is missing. Next we calculate the user interest truncation factor, which indicates the number of interactions that should be generated on each missing time pivot:
\begin{equation}\label{8}
    l_{num} = max(1, \frac{|t^o|}{|t^p_{set}|}).
\end{equation}

\noindent \textbf{Augmented User graphs Inference Process.} Afterwards, we sample through the trained generator by utilizing the latent factor to infer potential user preferences. We proceed with $K$ sampling steps and inference to derive the augmented adjacency matrix $\hat{\boldsymbol{z}}_0^{t} = \boldsymbol{\mu}_{\theta}(\boldsymbol{z}_{K}^t,\mathbf{c}_{t}, K)$ at a given time $t$. The augmented adjacency matrix predicts potential interactions within the user graph. Note that, as proven in previous work \cite{wang2023diffusion,jiang2024diffkg}, the inference step is optimal when $K$ is set to 0, in this case $\boldsymbol{z}_{K}^t = \boldsymbol{z}_{0}^t$. With this setting, the process is considered as denoising the sparse recommendation data.

With the set of uncovered pivots $t^p_{set}$ and the user interest truncation factor $l_{num}$ obtained from the above calculation, we identify the top $l_{num}$ highest scoring probabilistic interaction edges. By merging these edge sets with the original graph, we form the augmented graph $\mathbf{\mathcal{G}}_{aug}$, which can be expressed as
\begin{equation}\label{9}
\mathbf{\mathcal{G}}_{us}^{aug} = \mathbf{\mathcal{G}} \cup \text{topK}(\mathbf{\mathcal{G}}_{us}^{t_1}) \cup \text{topK}(\mathbf{\mathcal{G}}_{us}^{t_2}) \cup \ldots \cup \text{topK}(\mathbf{\mathcal{G}}_{us}^{t_m}).
\end{equation}
Through user preference truncation factors, we infer the potential user interests in the missing temporal pivots. With this inference process, we obtain enhanced user time graphs, which enhance the continuous dynamic interests of users on the timeline.

\subsubsection{\textbf{Generalized Graph Neural ODEs}}
\ 
\newline
The time-guided diffusion model interpolates the edges of user graph structures with extreme temporal imbalances to maintain the presence of nodes at pivot times. With this strategy, we enhance the intrinsic factors that influence user preferences. However, in long-term time user interest is simultaneously influenced by objective extrinsic factors. In this case, user interest changes may not match the dynamic item distributions. To capture the temporal consistency of users and items during evolution, we propose a generalized graph neural ODE model. It can model time-synchronous user and item dynamics.

\noindent \textbf{Time Sensitive Latent State Encoder.} To efficiently capture temporal information, we propose an attention-based GNN as our graph encoder. Specifically, given the initialized item embedding $\mathbf{x}$ and two graphs $\{\mathbf{\mathcal{G}}^{aug}_{us}, \mathbf{\mathcal{G}}_{cs}\}$, the item representations of both graphs are encoded as follows:
\begin{equation}\label{10}
\begin{aligned}
         \alpha_{ij} = \text{Sigmod}\left( \mathbf{a}^\top \left[ \mathbf{W}_Q \mathbf{x}_i, \mathbf{W}_K \mathbf{x}_j,  \Phi(t) \right] \right) , \\
         \mathbf{e}_{us} =  \sum_{j \in \mathcal{N}_i^{us}} \alpha_{ij} \mathbf{W}_V \mathbf{x}_j + \mathbf{W}_l \mathbf{x}_i,  \ \mathbf{e}_{cs} =\sum_{j \in \mathcal{N}_i^{cs}} \alpha_{ij} \mathbf{W}_V &\mathbf{x}_j + \mathbf{W}_l \mathbf{x}_i,
\end{aligned}
\end{equation}
where $\alpha_{ij}$ represents the attention weight of node $i$ towards node $j$, $\mathbf{W}_Q$, $\mathbf{W}_K$, $\mathbf{W}_V$, and $\mathbf{W}_l$ denote the weight matrices, $\Phi(t)$ signifies the time position embedding, and $\mathcal{N}_i$ denotes the set of neighbors of node $i$. Through the application of a graph encoder, the initialized item embedding $\mathbf{x}$ is transformed into two item representations $\mathbf{e}_{us}$ and $\mathbf{e}_{cs}$, which tend to focus on the changes in the users' interests and the evolution of item distributions, respectively. 

\noindent \textbf{Generalized Graph ODE Solver Function.} To accurately align with the dynamic evolution of item representations in continuous time, we propose a temporal graph ODE solver function = $\mathbf{f}_\theta$ to simulate the derivative in the ODE process. To predict the representation of items at any given moment, we encode the timestamps and corresponding graph structures in the ODE solver function, where $\mathbf{f}_\theta = \mathbf{f}_\theta \{ \mathbf{e}_{us}(t), \mathbf{e}_{cs}(t), \mathbf{g}(t), \mathbf{\mathcal{G}}^{aug}_{us}(t), \mathbf{\mathcal{G}}_{cs}(t)\}$ and $\mathbf{g}$ denotes the timestamp encoding function.

Upon obtaining the precise time embedding $\mathbf{g}(t)$ through position encoding, we align the item representations of the user and item graphs at time $t$ by means of the ODE function $\mathbf{f}_\theta$. Given the adjacency matrix $\mathbf{A}_{us} \in \mathbf{\mathcal{G}}^{aug}_{us}$ and the corresponding degree matrix $\mathbf{D}_{us}$, we opt for the normalized adjacency matrix $\tilde{\mathbf{A}}_{us}$ to facilitate neighborhood information transfer among graph nodes. Notably, as time progresses within the ODE, multiple layers are stacked, and normalization proves effective in mitigating potential gradient explosion issues. Given the directed nature of both graph types, we employ normalization via $\mathbf{D}_{us}^{-1} \mathbf{A}_{us}$ instead of $\mathbf{D}_{us}^{-\frac{1}{2}} \mathbf{A}_{us} \mathbf{D}_{us}^{-\frac{1}{2}}$. The ODE solver function is defined as follows:
\begin{equation}\label{11}
\begin{aligned}
& \frac{d \mathbf{\tilde{e}}^{t}}{dt} =  f_{us} (\mathbf{e}_{us}, g(t) ) + f_{cs} (\mathbf{e}_{cs}, g(t) ), \  f_{cs}: \tilde{\mathbf{e}}_{cs}^{t} = \mathbf{W}_a [\mathbf{e}_{cs}^{t}, \mathbf{g}(t)] \\
&f_{us}: \mathbf{\tilde{e}}_{us}^{l+1(t)} = \mathbf{W}_b [\mathbf{e}_{us}^{l(t)}, \mathbf{g}(t)], \
\mathbf{e}_{us}^{l+1(t)} = \sigma\left(\tilde{A}_{us}^{t} \mathbf{e}_{us}^{l(t)} \mathbf{W}_c\right)+\mathbf{e}_{us}^{l(t)} 
\end{aligned}
\end{equation}
where $\mathbf{W}_a$, $\mathbf{W}_b$, $\mathbf{W}_c$ denote trainable parameter matrices, $\sigma$ represents the activation function, $l$ indicates the number of layers, and $\tilde{\mathbf{A}}_{us}^{t}$ signifies the normalized adjacency matrix of the subgraph $\mathbf{\mathcal{G}}_{us, <t}^{aug}$. The derivatives of item representations at $t$ timestamps are modeled by combining two networks $f_{us}$ and $f_{cs}$. As the user time graph is augmented by a time-guided diffusion generator, we meticulously mine higher-order neighborhood relationships in the $f_{us}$. In contrast, the item evolution graph contains the global distribution of items, and we use the MLP to explicitly model the influence of external factors within the $f_{cs}$. During the ODE evolution, both factors simultaneously affect the item state at the corresponding moment. Ultimately, we use the Runge-Kutta method to obtain the final item representation $\tilde{\mathbf{e}}_{us}^{t_{n+1}}$ and $\tilde{\mathbf{e}}_{cs}^{t_{n+1}}$ at the target time $t_{n+1}$.

\subsection{Model Prediction and Optimization}

By evolving the states of both graphs through a neural ODE process, we are able to obtain the target time item representations $\mathbf{\tilde{e}}_{us}^{t_{n+1}}$ and $\mathbf{\tilde{e}}_{cs}^{t_{n+1}} $ at the target time. The final item vectors are then fed through a transformer decoder \cite{vaswani2017attention}, and obtain the final sequence representation:
\begin{equation}\label{12}
\mathbf{h}_s  = ||Decoder(\mathbf{\tilde{e}}_{us}^{t_{n+1}})|| + ||Decoder(\mathbf{\tilde{e}}_{cs}^{t_{n+1}})||.  
\end{equation}
where $||\cdot||$ is the $L_2$ norm.

To predict the probability of user sequence interaction with a specific item $v$, we compute the dot product between users' sequence embedding $\mathbf{h}_s$ and the target item embedding $\mathbf{\Tilde{e}}_v$, in order to assess their correlation:
\begin{equation}\label{13}
    \hat{\mathbf{y}} = \text{softmax}({\lVert {\mathbf{h}_s} \rVert}^\intercal {\lVert \mathbf{\Tilde{e}}_v\rVert}).
\end{equation}

We employ the cross-entropy loss function to calculate the optimized objective for the primary recommendation task, the definition is as follows:

\begin{equation}\label{14}
    \mathcal{L}_{rec} = -\sum_{v=1}^{\mid V\mid}{{\mathbf{y}}}\log(\hat {\mathbf{y}})
+(1-{\mathbf{y}})\log(1-\hat {\mathbf{y}}).
\end{equation}

Note that we alternately train the two modules, time-guided diffusion generation and generalized graph neural ODEs, to iteratively improve the recommendation performance. The specific training process is shown in the Appendix \ref{app:algorithm}. In addition, we provide a discussion \ref{tab:discussion} that further elaborates on the effectiveness of our iterative update strategy.

\subsection{Discussion }
\label{tab:discussion}

In this section, we examine the effectiveness and necessity of our iterative updating strategy. 

First, directly applying a diffusion model is not feasible. Our task requires predicting user preferences at a target time, which involves understanding continuous time in long sequences. However, as user interests tend to drift over time, diffusion models struggle to capture accurate distributions over extended periods, as evidenced by the poor performance of two diffusion models in Table \ref{table:baseline}. This observation highlights the need to integrate representations with accurate temporal information into the diffusion process. 

Second, a continuous-time representation using a graph neural ODE is essential. Although our generator is trained on existing timestamps, it must also reason about representations at other times. Neural ODEs can model node representations at arbitrary time points, providing a foundation for the diffusion model to learn feature distributions beyond the training timestamps. 

Finally, the diffusion model and the graph neural ODE are complementary. The diffusion generator enhances the graph neural ODE by supplying augmented graphs that capture detailed changes in user interests. While the neural ODE offers continuous-time representations that help the diffusion generator reason about distributions at any time. Together, these components reinforce each other, leading to improved overall performance.

\section{EXPERIMENTS}

We conduct extensive experiments to verify the performance of our TGODE model by answering the following questions: 

\begin{itemize}
\item {\bfseries RQ1: }How does the performance of our TGODE compare to different types of recommendation methods in competitive scenarios?
\item {\bfseries RQ2: }What is the impact of each specific key module within our TGODE framework on the overall performance?
\item {\bfseries RQ3: }How does our method perform with different sequence decoders?
\item {\bfseries RQ4: }How does TGODE perform in handling datasets with different levels of sparsity?
\item {\bfseries RQ5: }How interpretable is TGODE in real situations?
\end{itemize}

\subsection{Experiment Settings}
\subsubsection{Dataset and Evaluation}
To verify the validity of our approach, we conduct extensive experiments on five public datasets: (a) Beauty, (b) Sports, (c) Toys, (d) Video, and (e) ML-100k, which are widely used in the SR tasks. The first four datasets come from the Amazon platform and ML-100k is obtained from the MovieLens. Unlike general SR tasks, we employ a distinct evaluation method that is more consistent with prediction under the target time. More details are shown in the Appendix \ref{app:dataset}.

\subsubsection{Baselines and Evaluation Metrics}
We evaluate our TGODE with different types of baseline comparisons: the traditional methods (NARM \cite{li2017neural}, SRGNN \cite{wu2019session}, GRU4REC \cite{hidasi2015session}); the transformer methods (SASRec \cite{kang2018self}, SSE-PT \cite{wu2020sse}, CORE \cite{hou2022core}, MAERec \cite{ye2023graph}); the diffusion methods (DreamRec, DiffRec); and the continues time methods (TisasRec \cite{li2020time}, GNG-ODE \cite{guo2022evolutionary}, GDERec \cite{qin2024learning}). We adopt three widely used evaluation metrics in recommendation: Recall@$k$ (R@$k$, $k$=5, 10, 20), MRR@$k$ (M@$k$, $k$=5, 10, 20) and NDCG@$k$ (N@$k$, $k$=5, 10, 20).

\subsubsection{Experiment Details}
We implement our proposed method using PyTorch and optimize the parameters with the Adam optimizer. Since the original user graph is too sparse, we set the step $K$ to 5 in the training phase and set the step $K$ to 0 in the generation process referring to \cite{wang2023diffusion,jiang2024diffkg}. The hyperparameters for which we performed the grid search are shown below: the learning rate is sampled between 1e-1 and 1e-5. The range of the number of layers of the graph neural network is \{1, 2, 3, 4, 5\} and the number of heads of the Transformer ranges from 1 to 3. The dimension of embeddings varies in the range of \{32, 64, 128, 256\}. After we search for the optimal hyperparameters, we compare the results with other baseline models under the same settings. 

\begin{table*}[t]
\caption{Performance comparison among TGODE and twelve baselines over five datasets.}
\vspace{-5pt}
\scalebox{0.7}[0.7]{
  \centering

  \begin{tabular}{c|c|ccc|cccc|cc|ccc|c|c}
    \toprule
    \multirow{2}{*}{Dataset} & \multirow{2}{*}{Metrics} &\multicolumn{3}{|c|}{Tranditional Method} & \multicolumn{4}{|c|}{Transformer Method} & \multicolumn{2}{|c|}{Diffusion Method} & \multicolumn{3}{|c|}{Continues Time Method} & Ours & \multirow{2}{*}{\textit{improve.}} \\

     &  & NARM  & SRGNN & GRU4REC & SASRec & SSE-PT & CORE & MAERec & DreamRec & DiffRec & TisasRec & GNG-ODE & GDERec   & TGODE &   \\

    \hline
    \multirow{6}{*}{Beauty}

& R@5 & 0.0112 & 0.0015 & 0.0131 & 0.0223 & 0.0258 & 0.0263 & 0.0253 & 0.0005 & 0.0101 & 0.0145 & \underline{0.0304} & 0.0151 & \textbf{0.0422} & 38.82\% \\
& R@10 & 0.018 & 0.004 & 0.0223 & 0.0357 & 0.0443 & \underline{0.0541} & 0.0473 & 0.001 & 0.0181 & 0.0255 & 0.05 & 0.0267 & \textbf{0.0709} & 31.05\% \\
& R@20 & 0.0281 & 0.0083 & 0.0345 & 0.0529 & 0.0706 & \underline{0.09} & 0.0796 & 0.0018 & 0.0305 & 0.0435 & 0.0775 & 0.0436 & \textbf{0.1077} & 19.67\% \\
& N@5 & 0.0071 & 0.0007 & 0.0082 & 0.012 & 0.0154 & 0.0124 & 0.0149 & 0.0003 & 0.006 & 0.0085 & \underline{0.0192} & 0.0093 & \textbf{0.0243} & 26.56\% \\
& N@10 & 0.0093 & 0.0015 & 0.0111 & 0.0164 & 0.0214 & 0.0214 & 0.0219 & 0.0005 & 0.0085 & 0.012 & \underline{0.0255} & 0.0131 & \textbf{0.0336} & 31.76\%\\
& N@20 & 0.0118 & 0.0025 & 0.0142 & 0.0207 & 0.028 & 0.0305 & 0.03 & 0.0007 & 0.0117 & 0.0166 & \underline{0.0325} & 0.0173 & \textbf{0.0429} & 32.00\% \\

    \hline
    \multirow{6}{*}{Sports} 

& R@5 & 0.01 & 0.0112 & 0.0113 & 0.0101 & 0.0153 & 0.0164 & 0.0177 & 0.0003 & 0.0098 & 0.0099 & \underline{0.0188} & 0.013 & \textbf{0.0257} & 36.70\%
\\
& R@10 & 0.0159 & 0.0176 & 0.0155 & 0.0163 & 0.0267 & \underline{0.0323} & 0.0296 & 0.0006 & 0.0162 & 0.0161 & 0.0319 & 0.0211 & \textbf{0.0426} & 31.89\%
\\
& R@20 & 0.026 & 0.0282 & 0.0261 & 0.0263 & 0.0435 & \underline{0.0552} & 0.0468 & 0.0009 & 0.0259 & 0.0254 & 0.0504 & 0.0348 & \textbf{0.0664} & 20.29\%
\\
& N@5 & 0.0067 & 0.0073 & 0.0073 & 0.0054 & 0.0095 & 0.0078 & 0.0114 & 0.0002 & 0.0059 & 0.0063 & \underline{0.0119} & 0.0083 & \textbf{0.0152} & 27.73\%
\\
& N@10 & 0.0086 & 0.0094 & 0.0086 & 0.0074 & 0.0132 & 0.0129 & 0.0152 & 0.0002 & 0.008 & 0.0083 & \underline{0.0161} & 0.0109 & \textbf{0.0206} & 27.95\%
\\
& N@20 & 0.0111 & 0.012 & 0.0113 & 0.0099 & 0.0174 & 0.0187 & 0.0196 & 0.0003 & 0.0104 & 0.0107 & \underline{0.0208} & 0.0144 & \textbf{0.0266} & 27.88\%
\\

    \hline
    \multirow{6}{*}{Toys} 
    
& R@5 & 0.0082 & 0.004 & 0.0079 & 0.0168 & 0.0174 & \underline{0.0283} & 0.0176 & 0.0007 & 0.0063 & 0.0127 & 0.0184 & 0.0071 & \textbf{0.0392} & 38.52\%
\\
& R@10 & 0.0107 & 0.0084 & 0.0123 & 0.0232 & 0.0297 & \underline{0.047} & 0.0307 & 0.0012 & 0.0113 & 0.0186 & 0.0287 & 0.0131 & \textbf{0.0620} & 31.91\%
\\
& R@20 & 0.0149 & 0.0131 & 0.0209 & 0.0301 & 0.0456 & \underline{0.0726} & 0.0497 & 0.002 & 0.0182 & 0.0262 & 0.0446 & 0.0226 & \textbf{0.0869} & 19.70\%
\\
& N@5 & 0.0056 & 0.0024 & 0.005 & 0.0094 & 0.0095 & \underline{0.0162} & 0.0102 & 0.0004 & 0.0034 & 0.0074 & 0.0118 & 0.0042 & \textbf{0.0221} & 36.42\%
\\
& N@10 & 0.0064 & 0.0038 & 0.0064 & 0.0115 & 0.0135 & \underline{0.0199} & 0.0144 & 0.0006 & 0.005 & 0.0093 & 0.0151 & 0.0061 & \textbf{0.0291} &  46.23\%
\\
& N@20 & 0.0075 & 0.005 & 0.0086 & 0.0132 & 0.0175 & \underline{0.0254} & 0.0192 & 0.0008 & 0.0067 & 0.0113 & 0.0191 & 0.0085 & \textbf{0.0354} & 39.37\%\\

    \hline
    \multirow{6}{*}{Video}

& R@5 & 0.0145 & 0.014 & 0.0211 & 0.0301 & 0.0335 & 0.0338 & \underline{0.0465} & 0.0004 & 0.0155 & 0.0217 & 0.0416 & 0.0186 & \textbf{0.0565} & 21.51\%
\\
& R@10 & 0.0247 & 0.0246 & 0.0354 & 0.0508 & 0.0602 & 0.0707 & \underline{0.0775} & 0.0005 & 0.0271 & 0.0351 & 0.0678 & 0.0333 & \textbf{0.0930} & 20.00\%
\\
& R@20 & 0.0375 & 0.0342 & 0.0564 & 0.0795 & 0.0981 & 0.1224 & \underline{0.1238} & 0.001 & 0.046 & 0.0559 & 0.107 & 0.0569 & \textbf{0.1423} & 14.94\%
\\
& N@5 & 0.0099 & 0.0096 & 0.0136 & 0.0159 & 0.0207 & 0.0169 & \underline{0.0284} & 0.0004 & 0.0088 & 0.0139 & 0.0261 & 0.0116 & \textbf{0.0341} & 20.07\%
\\
& N@10 & 0.0132 & 0.0131 & 0.0182 & 0.0225 & 0.0293 & 0.0287 & \underline{0.0383} & 0.0004 & 0.0125 & 0.0182 & 0.0346 & 0.0163 & \textbf{0.0462} & 20.63\%
\\
& N@20 & 0.0165 & 0.0154 & 0.0235 & 0.0298 & 0.0388 & 0.0418 & \underline{0.0501} & 0.0005 & 0.0173 & 0.0235 & 0.0446 & 0.0222 & \textbf{0.0588} & 17.37\%\\

    \hline
    \multirow{6}{*}{ML-100K} 
 
& R@5 & 0.0068 & 0.0077 & 0.0128 & 0.0144 & 0.022 & 0.0142 & 0.0206 & 0.0035 & 0.0035 & 0.0087 & \underline{0.0388} & 0.013 & \textbf{0.0433} & 11.60\%
\\
& R@10 & 0.0133 & 0.0109 & 0.0242 & 0.0266 & 0.0436 & 0.0313 & 0.0399 & 0.0072 & 0.0091 & 0.0191 & \underline{0.0684} & 0.0275 & \textbf{0.0831} & 21.49\%
\\
& R@20 & 0.0241 & 0.0208 & 0.0451 & 0.0475 & 0.0819 & 0.0637 & 0.0771 & 0.015 & 0.0228 & 0.0377 & \underline{0.1188} & 0.0524 & \textbf{0.1486} & 25.08\%
\\
& N@5 & 0.0043 & 0.0046 & 0.0078 & 0.0085 & 0.0136 & 0.0074 & 0.012 & 0.0022 & 0.0021 & 0.0054 & \underline{0.0245} & 0.0101 & \textbf{0.0271} & 10.61\%
\\
& N@10 & 0.0065 & 0.0056 & 0.0125 & 0.0125 & 0.0209 & 0.0133 & 0.0184 & 0.0034 & 0.0041 & 0.0089 & \underline{0.0347} & 0.0155 & \textbf{0.0413} & 19.02\%
\\
& N@20 & 0.0094 & 0.0084 & 0.0177 & 0.0181 & 0.0312 & 0.022 & 0.0282 & 0.0054 & 0.0078 & 0.0138 & \underline{0.0479} & 0.0226 & \textbf{0.0586} & 22.34\%\\

    \bottomrule
  \end{tabular}
  \vspace{-15pt}
  \label{table:baseline}
}
\end{table*}
  \vspace{-4pt}

\subsection{Comparison of Performance (RQ1)}

In this section, we conduct a comprehensive performance evaluation of TGODE and the aforementioned baselines. For simplicity, here we present the experimental results for Recall@$k$ and NDCG@$k$, while the results for MRR@$k$ can be found in the Appendix \ref{app:baselines}. Table \ref{table:baseline} presents the experimental results of each model on three datasets, yielding the following observations:

The recommendation models utilizing Transformers, such as SASRec, SSE-PT, CORE, and MAERec, demonstrate superior performance on multiple datasets compared to traditional models like NARM, SRGNN, and GRU4Rec. These observations highlight the practicality of utilizing Transformers for modeling sequential information and their adaptability to various recommendation tasks in different contexts.

Models that consider continuous time, such as TisasRec, GNG-ODE, and GDERec, have a competitive advantage over the two types of models mentioned above. This emphasizes the importance of capturing the dynamically changing aspects of user preferences. However, they ignore the temporal sparsity of user interests. As a result, the user preferences captured by these models change in time in jumps, leading to biased recommendation results.

Approaches based on diffusion models, such as DreamRec and DiffRec, perform poorly on SR tasks. In short-term historical interactions, diffusion models are able to generate appropriate recommended items with their denoising ability. However, under long-term historical interactions, users' interests change dynamically. At this point, the diffusion model may generate previously preferred items that do not match the current user preferences. Therefore, it is inappropriate to simply apply the diffusion model in long sequence recommendation.

Our approach consistently outperforms the baselines on multiple metrics, effectively addressing the challenges of dynamic user preferences and uneven item distribution. We mine temporally-guided diffusion models for temporal sequences of user interaction information and use user preference inference to generate latent user interests under missing temporal pivots. Afterwards, the graph neural ODE model matches the consistency of user interests and item distributions over the evolutionary process, effectively improving the recommendation performance.

\vspace{-2pt}
\subsection{Ablation Experiment (RQ2)}

The ablation experiment section compares three model variants to assess the impact of each module in TGODE on performance.

\begin{itemize}
\item The {\bfseries base} variant removes all valid modules.
\item The {\bfseries w/o Diff} variant removes the diffusion generator module.
\item The {\bfseries w/o ODE} variant removes the generalized graph neural ODEs module.
\item The {\bfseries w/o cs} variant removes the item evolution graph.

\end{itemize}
\begin{table}[t] 
 
\caption{Ablation study.}
\scalebox{0.85}[0.85]{
	\centering
	\begin{tabular}{cc|cc|cc|cc} 
	\toprule
	\multicolumn{2}{c|}{Ablation}& \multicolumn{2}{c|}{Beauty}& \multicolumn{2}{c|}{Sports} & \multicolumn{2}{c}{Toys} \\
        \cline{3-8}
	\multicolumn{2}{c|}{Settings}&R@20&N@20&R@20&N@20&R@20&N@20\\  
	\hline 
	\hline
	\multicolumn{2}{c|}{TGODE}& \textbf{0.1077} & \textbf{0.0429} & \textbf{0.0664} &\textbf{0.0266} & \textbf{0.0869} & \textbf{0.0354} \\   
        \hline
        \multicolumn{2}{c|}{base}& 0.0594 &  0.0235 & 0.0448 & 0.0184 & 0.0189 & 0.0077 \\
        \multicolumn{2}{c|}{w/o Diff}& 0.1008 & 0.0393 & 0.0609 & 0.0233 & 0.0846 & 0.0339 \\
        \multicolumn{2}{c|}{w/o ODE}& 0.0733 & 0.0294 & 0.0527 & 0.0220 & 0.0476 & 0.0191 \\
        \multicolumn{2}{c|}{w/o cs}& 0.1023 & 0.0400 &  0.0557 & 0.0224 &  0.0803 & 0.0323 \\
	\bottomrule 

	\end{tabular}
\vspace{-15pt}
 
 \label{table:ablation}
 }
 
\end{table}
  \vspace{-4pt}

\begin{figure}[t]
    \centering
    \includegraphics[width=\linewidth]{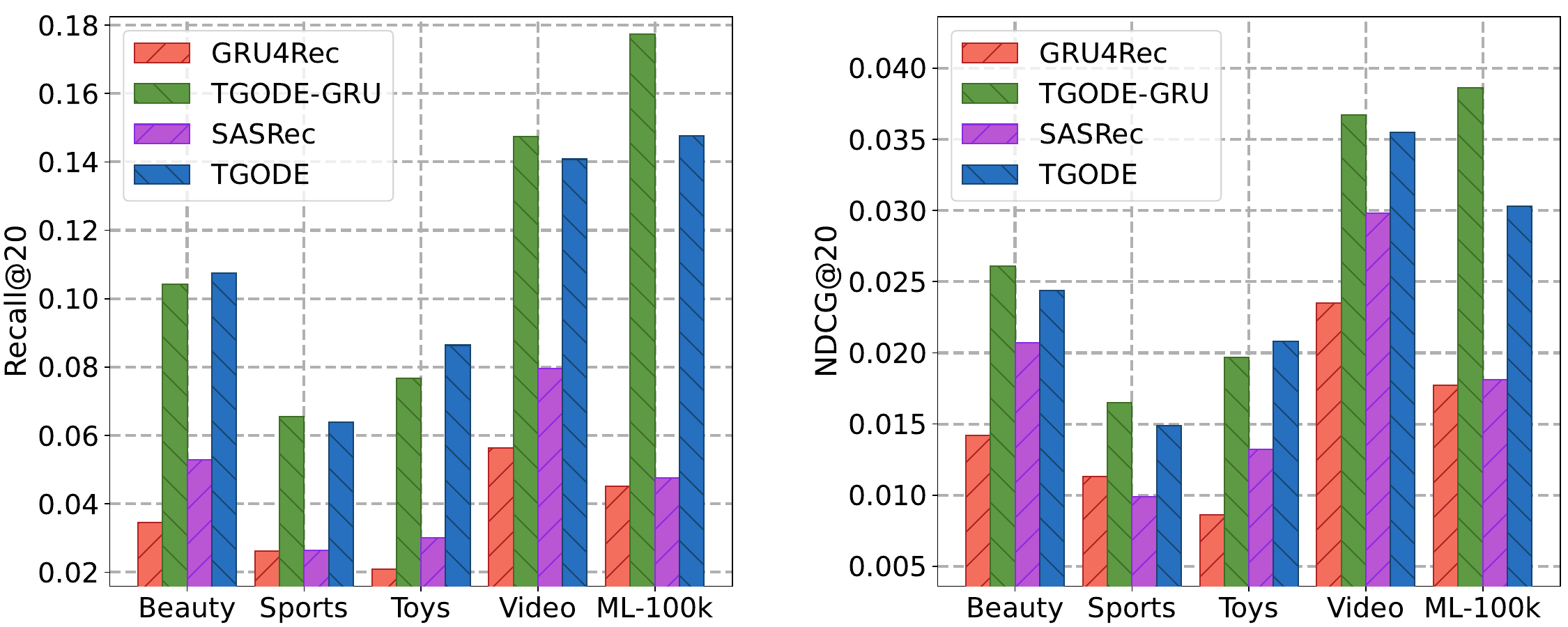}
    
    \caption{Comparison with different sequence decoders.}
    \vspace{-15pt}
    \label{fig:comparision}
\end{figure}

Table \ref{table:ablation} shows the performance comparison of our original model and its four ablated variants across two evaluation metrics. Based on the ablation results, the following observations can be analyzed and obtained:

From the significant performance degradation observed in the variant "base" compared to the original model, we can observe that merely utilizing a Transformer for sequence encoding in recommendation tasks is insufficient. 
This variant ignores irregular user interests and highly uneven item distributions, leading to great performance degradation.

From the experimental results of the variant "w/o Diff", it is evidenced for the efficacy of the diffusion model in reconstructing interaction sequences, consequently mitigating irregular user interests. With our designed time-guided strategy and user preference inference, the generator is able to consciously fill in user interactions in irregular temporal pivots, thus enhancing continuous user interest changes.

From the experimental results of the variant "w/o ODE", we can detect that ODE effectively captures and aligns dynamic user preferences and item distributions. The performance variations across datasets highlight the significance of this variant, with a notable 45\% performance decrease observed on the Toys dataset.

The experimental results of the variant "w/o cs" elucidate the impact of item distributions bias on the unbiased representation of the model. By neglecting the temporal popularity information of items, the model is unable to take into account changes and trends over time in items influenced by external factors. Specifically, the user's current purchasing behavior may simply be influenced by the most recently popular items, rather than his personal preferences.
\vspace{-4pt}
\subsection{Performance with Different Decoders (RQ3)}
To examine the expressive capacity of different encoders for sequential representation and their impact on the overall model performance, we employ two encoding approaches: GRU and Transformer. Additionally, we apply GRU4Rec (GRU) and SASRec (Transformer) as a comparison.

Figure \ref{fig:comparision} illustrates the impact results of our model on two different sequence encoding modules related to two baselines that utilize the corresponding module. We derive two observations from the results. Firstly, SASRec outperforms GRU4Rec due to the utilization of the Transformer framework, which enables the modeling of global contextual structures. In contrast, GRU4Rec's performance is limited as it relies solely on modeling long-term temporal dependencies, and its encoding capacity is constrained by its relatively simple sequential structure. This is also why we outperform the GRU variant in Beauty and Toys data. Secondly, GRU4Rec achieves performance close to SASRec on the Sports and ML-100K datasets. This can be attributed to the fragmented nature of interactions in the datasets, where there is no pronounced sequential order but rather a certain degree of long-term dependency. Consequently, our variant of GRU exhibits better performance in such scenarios. Overall, both variants of our approach outperform the direct application of the two corresponding sequence decoders, demonstrating the advancement of our proposed model.

\begin{figure}[t]
    \centering
    \includegraphics[width=\linewidth]{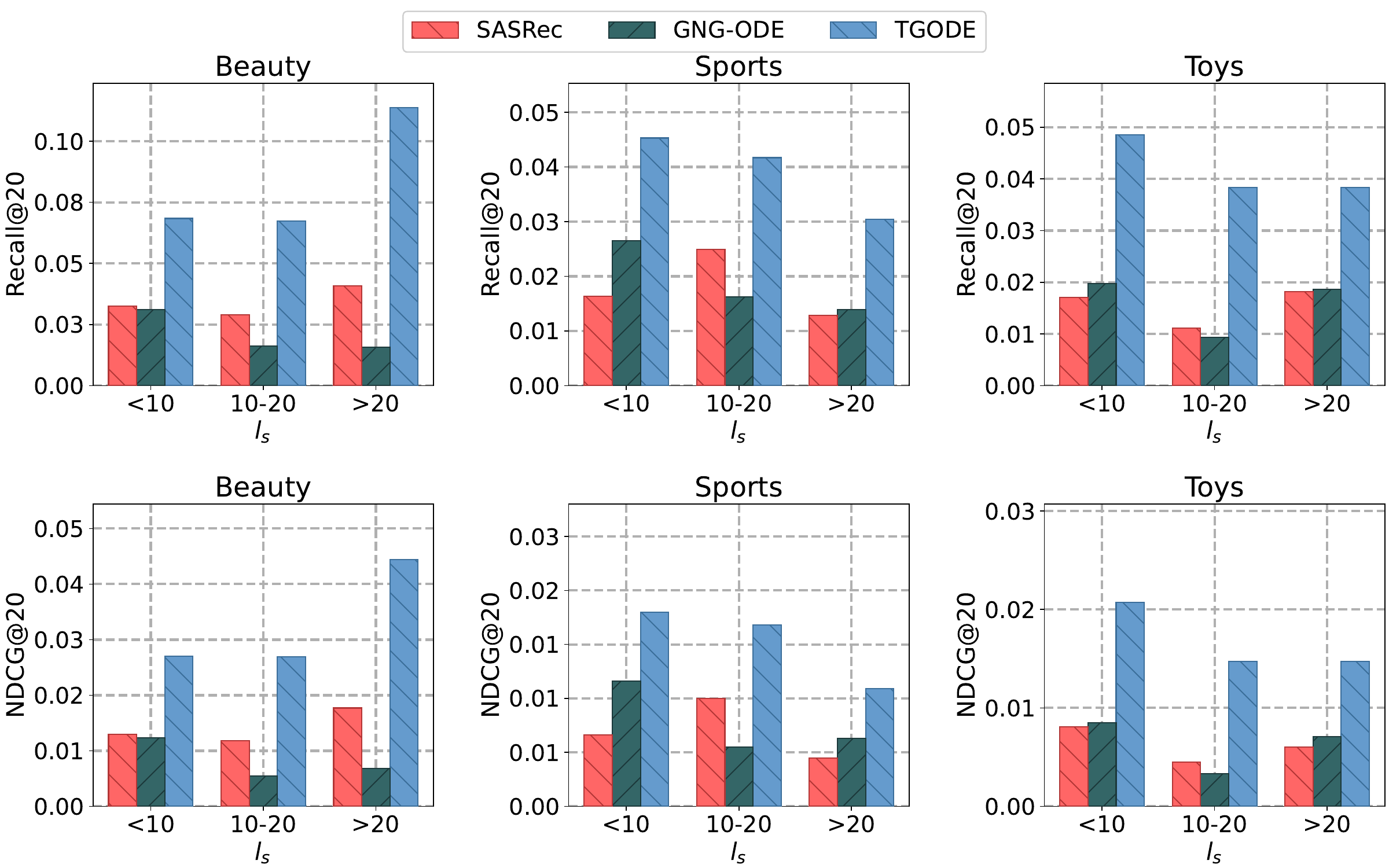}

    \caption{Sequence length within <10, 10-20, >20.}
\vspace{-15pt}
    \label{fig:seqlen}
\end{figure}
\vspace{-4pt}

\subsection{In-depth Exploration of Model (RQ4)}

In real-world datasets, sequence length is indicative of the sustained preferences among diverse user populations. To explore the efficacy of our TGODE model in capturing sustained preferences across different interest groups, we partition the sequence lengths into three categories: <10, 10-20, and >20. We compare the performance of our TGODE model with two baseline models (SASRec and GNG-ODE) as shown in Figure \ref{fig:seqlen}.

From the illustration, we can obtain three key observations. Firstly, when sequences are short, user interactions tend to be sparse with cold start scenarios, and there is little need for long-term memory or complex modeling of sequential dependencies as the contextual information is relatively limited. This allows the SASRec model, which utilizes the Transformer architecture, to more effectively model and utilize the limited contextual information when handling short sequences. Secondly, irregular time intervals within longer sequences will present challenges for the Transformer architecture in encoding long-range dependencies. In contrast, the GNG-ODE model excels at capturing such dependencies by modeling the continuous evolution process of sequences. Notably, in the Beauty dataset, SASRec outperforms GNG-ODE due to relatively shorter sequence lengths, which limit the manifestation of long-distance dependency relationships. Thirdly, our TGODE outperforms the two baselines across three variants of datasets with different sequence lengths. Our approach enhances temporally irregular user behavior, which enables capturing user preferences on sparse short sequences. At the same time, we match the patterns of evolution of user interests and item distributions over time, which allows us to improve the recommendation performance under processing long sequences.

\begin{figure}[t]
    \centering
    \includegraphics[width=\linewidth]{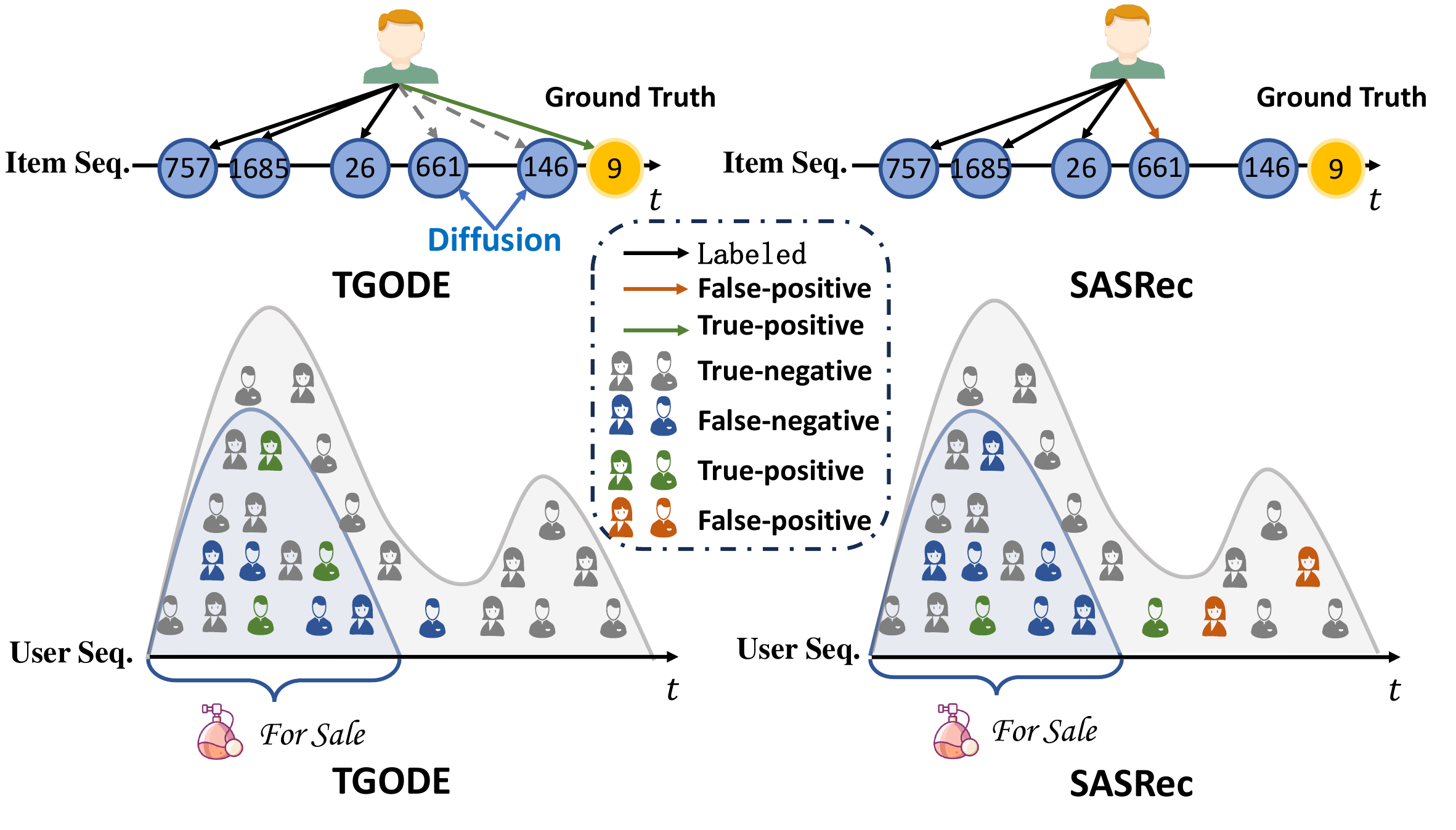}
    \caption{Case study of two examples on the Beauty dataset.}

    \label{fig:casestudy}
\vspace{-15pt}
\end{figure}
\vspace{-4pt}

\subsection{Case Study (RQ5)}
To validate TGODE's effectiveness, we extract two representative examples from the Beauty dataset. We also compare them with SASRec in terms of item and user sequences to demonstrate our approach's efficacy.

In Figure \ref{fig:casestudy}, we extract a user's interaction item sequence, where temporal gaps between items are represented as time intervals. To better characterize distinct sample categories, we utilize patterns of varying colors to represent them. Specifically, green patterns are employed to denote True-positive instances, while other color schemes are as annotated in the figure legend. Labeled data $v_{757}$, $v_{6185}$, and $v_{26}$ are included, and the ground truth item to be predicted is $v_9$. Our TGODE effectively predicts item $v_9$ by incorporating item $v_{146}$ through diffusion and capturing long-term temporal dependencies using ODE. In contrast, SASRec's lack of awareness of time interval information and long-distance dependencies results in an erroneous prediction of item $v_{661}$ adjacent to the labeled data.

In the lower part of Figure \ref{fig:casestudy}, we present an example demonstrating the distribution of users' interactions with items over a specific period (shown in gray). We select a popular item (shown in blue) that receives loads of users' interactions during this period for illustration. Our TGODE capture the popularity of items, successfully predicts numerous user interactions during its peak popularity and decreases recommendations as its popularity declines. In contrast, SASRec lacks a sense of item popularity over time and continues recommending the item to users even when it is no longer popular.


\section{Conclusion}
In conclusion, this paper introduces TGODE, a novel framework that addresses the challenges of irregular user interests and uneven item distributions in sequential recommendation systems. By iteratively training the time-guided diffusion generator and generalized graph neural ODEs, TGODE captures dynamic user behaviors and shifting item trends. Extensive experiments on multiple real-world datasets confirm that TGODE significantly outperforms existing state-of-the-art methods, demonstrating its effectiveness in enhancing recommendation accuracy.

\bibliographystyle{ACM-Reference-Format}
\balance
\bibliography{main}

\appendix

\section{Appendix}
\vspace{-2pt}
\subsection{Algorithm Details}
\label{app:algorithm}
The training process for our TGODE model is shown in Algorithm \ref{al}.
\vspace{-2pt}
\begin{algorithm}[htbp]
	\caption{The Procedure of TGODE}
	\label{al}
        \renewcommand{\algorithmicrequire}{\textbf{Input:}} 
        \renewcommand{\algorithmicensure}{\textbf{Output:}} 
	\begin{algorithmic}[1]
        \REQUIRE The sequences $\mathcal{S}$ and initial item embedding $\textbf{x}$.
		\STATE Initialization parameters $\theta_{diff}$ and $\theta_{rec}$;
		\STATE Construct user time graph $\mathcal{G}_{us}$ and item evolution graph $\mathcal{G}_{cs}$;
        \FOR {each iteration} 
                 
            \FOR {each batch} 
                \STATE Get the sequence representation $\mathbf{h}_s^t$ under timestamp $t$ from $\theta_{rec}$;
                \STATE Calculate the initial input $\mathbf{z}^t_0$ and time embedding $\textbf{c}_t$ corresponding to the timestamp $t$ according to Equation (\ref{0}-\ref{1});
                \STATE Perform the Forward Process based on Equation (\ref{2}) and the Reverse Process based on Equation (\ref{3});
                \STATE Calculate the loss function $\mathcal{L}_{diff}$ of the time-guided diffusion module based on Equation (\ref{4}-\ref{6});
                \STATE Update parameters $\theta_{diff}$ to minimize $\mathcal{L}_{diff}$;
            \ENDFOR

            \STATE Calculate the the set of uncovered pivots $t^p_{set}$ and the user interest truncation factor $l_{num}$ according to Equation (\ref{7}-\ref{8});
            \STATE Generate $l_{num}$ graphs on the uncovered pivots through the trained generator and merge them with the original user graph $\mathcal{G}_{us}$ into an augmented graph $\mathcal{G}^{aug}_{us}$ through Equation (\ref{9});
            \FOR {each batch}
                \STATE Calculate the item representations $\mathbf{e}_{us}$ and $\mathbf{e}_{cs}$ through the time sensitive latent state Encoder based on Equation (\ref{10});
                \STATE The final item representations $\tilde{\mathbf{e}}_{us}^{t_{n+1}}$ and $\tilde{\mathbf{e}}_{cs}^{t_{n+1}}$ are derived at the target time $t_{n+1}$ through a generalized graph neural ODE solver function based on Equation (\ref{11});
                \STATE Obtain the representation of the sequence $\mathbf{h}_s$ through Equation (\ref{12}) and compute the recommendation loss function $\mathcal{L}_{rec}$ based on Equation (\ref{13}-\ref{14}).
                \STATE Update parameters $\theta_{rec}$ to minimize $\mathcal{L}_{rec}$;
            \ENDFOR
        \ENDFOR
        \ENSURE  Optimized $\theta_{diff}$ and $\theta_{rec}$.
	\end{algorithmic}
\end{algorithm}
\vspace{-2pt}
\subsection{Efficiency Analysis}
\label{tab:Efficiency}
We analyze the time complexity of our two main modules: the diffusion generator and the graph neural ODE. For the time-guided diffusion generator, the complexity is $O(m×(K_1 \times n^2+K_2 \times n^2 \times d))$, where $m$ is the number of generation, $K_1$ and $K_2$ are the steps of forward process and reverse process, $n$ is the number of items, $d$ is the hidden size. Unlike general diffusion models, we perform $m$-times generation in the inference process. Since the number of generation times can be manually tuned to balance performance and efficiency, the added time overhead is acceptable. Additionally, The complexity of ODE is $O(k \times(l \times E \times d+n \times d^2))$. $k$ is the step size of ODE, $l$ is the number of layers and $E$ is the number of edges. Here, the number of edges is the primary factor affecting complexity. Overall, the time complexity of our modules is mainly determined by the graph size, but the computational overhead can be managed by tuning the relevant parameters as the graph size increases.

\vspace{-2pt}

\subsection{Dataset and Evaluation}
\label{app:dataset}
\subsubsection{Dataset Description}
In this paper, we conduct experiments on five typical public datasets and compare TGODE with other baselines. This website (https://jmcauley.ucsd.edu/data/amazon/) contains all the datasets. A brief description of these five datasets is given below:
\textbf{Beauty}, \textbf{Sports}, \textbf{Toys} and \textbf{Video} collect interaction records between users and items in different fields on Amazon respectively. Users also have certain differences in their interaction preferences for items in different fields. Meanwhile, \textbf{ML-100k} is a stable benchmark dataset that tracks the ratings of 100k movies on MovieLens. The detailed information of these five datasets is shown in Table \ref{datasets}.

\begin{table}[tbp]
 \caption{Statistics of datasets}
\scalebox{1}[1]{
  \centering
  \begin{tabular}{cccccc}
    \toprule
    Dataset    & User & Item   & Inter & Ave.Len & Density \\
    \midrule
    Beauty     & 22363 & 12101 & 198502 & 8.87 & 99.92\%          \\
    Sports     & 35598 & 18357 & 296337  &8.32  & 99.95\%          \\
    Toys & 19412 & 11924 & 167597 & 8.63 & 99.92\%      \\
    Video & 24303 & 10672 & 231780  & 9.53 & 99.91\%      \\
    ML-100k & 943 & 1682 & 100000  & 106.04 & 93.69\%        \\
    \bottomrule
  \end{tabular}\label{datasets}
  }
  
\end{table}

\subsubsection{Evaluation}
\label{app:Evaluation}
In a realistic scenario, the recommender system only has data that is prior to the current moment. Naturally, all training data should be earlier in time than the evaluation and test data. Traditional sequence evaluation uses the leave-one-out method to predict the last interaction of each user. In contrast, we divide all interactions in chronological order in a ratio of 8:1:1. With this division method, we consider all interactions as prediction targets. In this case, the input sequence is the sequence of corresponding user interactions previous to that target time.

\begin{figure}[htbp]
    \centering
    \subfigure[User Interaction Timelines ]{
	\centering
        \includegraphics[width=0.43\linewidth]{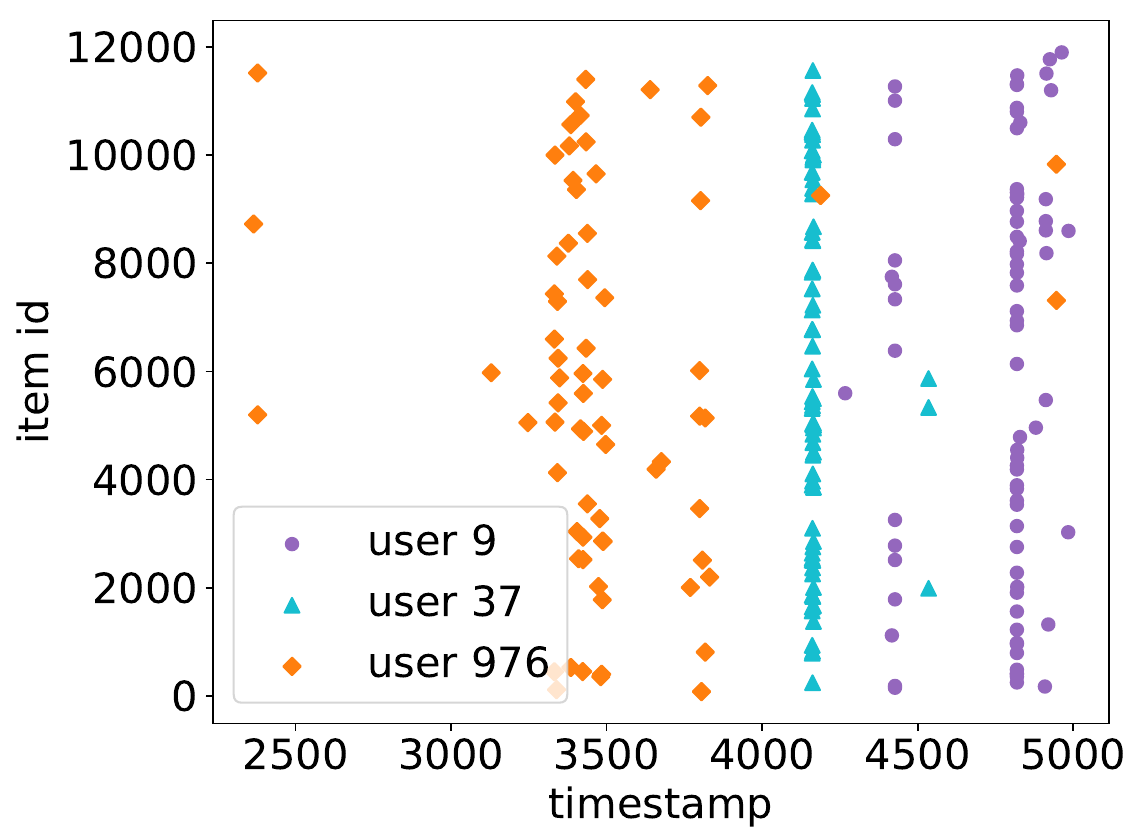}
        \label{t-a}
    }
    \centering
    \subfigure[Item Interaction Timelines  ]{
	\centering
        \includegraphics[width=0.43\linewidth]{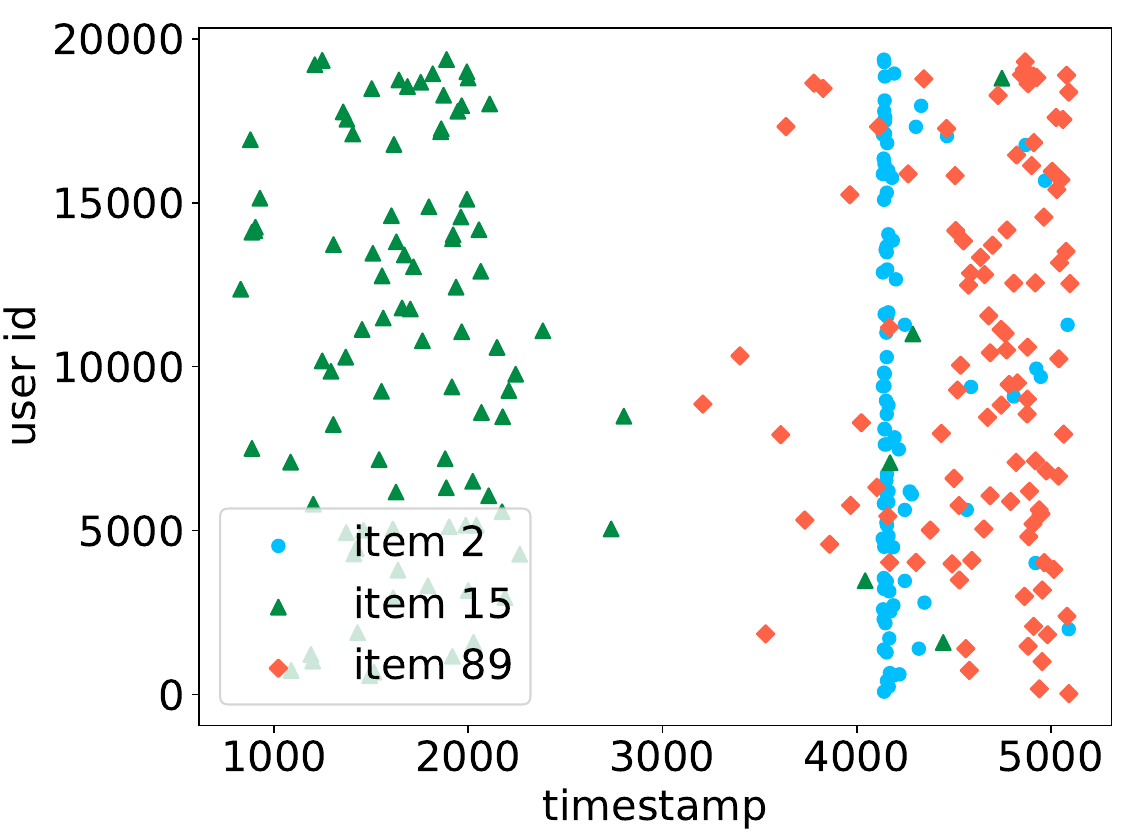}
        \label{t-c}
    }
    \centering
    \subfigure[Item Emergence Ratios ]{
	\centering
        \includegraphics[width=0.43\linewidth]{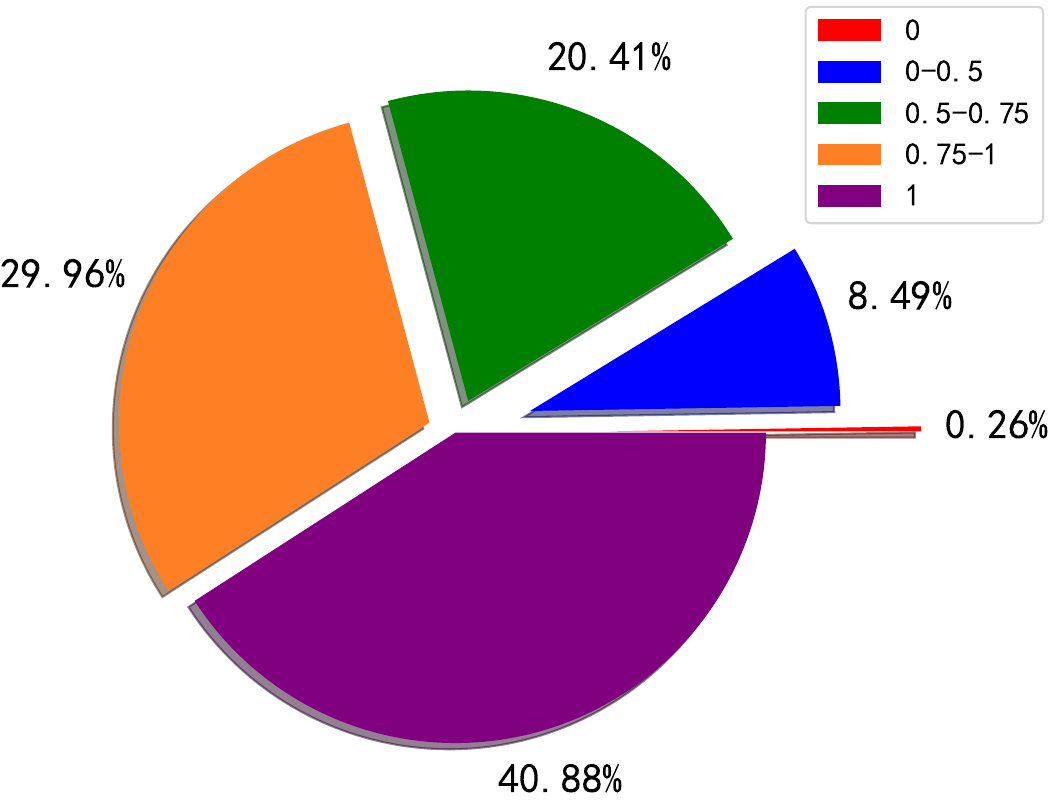}
        \label{t-b}
    }
    \vspace{-15pt}
    \caption{Data analysis regarding the Toys dataset.}
    \label{fig:toys}
\end{figure}

Similar to the approach in Section 2.2, we present the relevant analysis of the toys dataset in figure \ref{fig:toys}, including user interests, item emergence ratios, and item distributions.

As shown in figure \ref{t-a}, in the Toy dataset, the interactions of these users are also clustered within a certain time range, and exhibit irregular distribution at other timestamps.  
Additionally, as seen in figure \ref{t-b}, more than 80\% of the items have an occurrence rate higher than 75\%, while items with an occurrence rate below 10\% account for less than 10\%.
Furthermore, as shown in figure \ref{t-c}, item interactions with numerous users are also concentrated at specific timestamps.  
These analytical results exhibit the same trend as the Beauty dataset in Section 2.2, further emphasizing the necessity of considering irregular time intervals and modeling in a time-aware manner.
\vspace{-2pt}
\subsection{Data Analysis for Toys}
\label{app:DataAnalysis}
\vspace{-7pt}
\begin{table*}[t]
\caption{Additional comparison results with the baselines on the MRR metric.}
\vspace{-5pt}
\scalebox{0.7}[0.7]{
  \centering

  \begin{tabular}{c|c|ccc|cccc|cc|ccc|c|c}
    \toprule
    \multirow{2}{*}{Dataset} & \multirow{2}{*}{Metrics} &\multicolumn{3}{|c|}{Tranditional Method} & \multicolumn{4}{|c|}{Transformer Method} & \multicolumn{2}{|c|}{Diffusion Method} & \multicolumn{3}{|c|}{Continues Time Method} & Ours & \multirow{2}{*}{\textit{improve.}} \\
    
     &  & NARM  & SRGNN & GRU4REC & SASRec & SSE-PT & CORE & MAERec & DreamRec & DiffRec & TisasRec & GNG-ODE & GDERec   & TGODE &   \\

    \hline
    \multirow{3}{*}{Beauty}

& M@5 & 0.0057 & 0.0005 & 0.0066 & 0.0086 & 0.0119 & 0.008 & 0.0115 & 0.0003 & 0.0046 & 0.0066 & \underline{0.0156} & 0.0075 & \textbf{0.0181} & 16.03\%\\
& M@10 & 0.0066 & 0.0008 & 0.0078 & 0.0104 & 0.0144 & 0.0116 & 0.0143 & 0.0003 & 0.0057 & 0.008  & \underline{0.0182} & 0.009 & \textbf{0.0219} & 20.33\%\\
& M@20 & 0.0073 & 0.001 & 0.0086 & 0.0116 & 0.0162 & 0.0141 & 0.0165 & 0.0004 & 0.0065 & 0.0092 & \underline{0.02} & 0.0102 & \textbf{0.0224} & 12.00\%\\

    \hline
    \multirow{3}{*}{Sports}

& M@5 & 0.0056 & 0.006 & 0.006 & 0.0039 & 0.0076 & 0.005 & 0.0094 & 0.0001 & 0.0046 & 0.0052 & \underline{0.0097} & 0.0068 & \textbf{0.0118} & 21.65\%\\
& M@10 & 0.0064 & 0.0069 & 0.0065 & 0.0047 & 0.0091 & 0.0071 & 0.0109 & 0.0002 & 0.0055 & 0.006 & \underline{0.0114} & 0.0079 & \textbf{0.0140} &22.81\%\\
& M@20 & 0.0071 & 0.0076 & 0.0073 & 0.0053 & 0.0103 & 0.0087 & 0.0121 & 0.0002 & 0.0061 & 0.0066 & \underline{0.0126} & 0.0088 & \textbf{0.0156} & 23.81\%\\

    \hline
    \multirow{3}{*}{Toys}

& M@5 & 0.0047 & 0.0018 & 0.0041 & 0.0069 & 0.0069 & 0.0075 & 0.0078 & 0.0003 & 0.0025 & 0.0057 & \underline{0.0096} & 0.0033 & \textbf{0.0164} & 70.83\%\\
& M@10 & 0.0051 & 0.0024 & 0.0047 & 0.0078 & 0.0085 & 0.0106 & 0.0095 & 0.0004 & 0.0031 & 0.0065 & \underline{0.011} & 0.0041 & \textbf{0.0192} & 74.55\%\\
& M@20 & 0.0054 & 0.0027 & 0.0052 & 0.0082 & 0.0096 & 0.0123 & 0.0108 & 0.0004 & 0.0036 & 0.007 & \underline{0.012} & 0.0047 & \textbf{0.0210} & 75.00\%\\

    \hline
    \multirow{3}{*}{Video}

& M@5 & 0.0084 & 0.0081 & 0.0112 & 0.0112 & 0.0165 & 0.0114 & \underline{0.0224} & 0.0003 & 0.0066 & 0.0114 & 0.021 & 0.0093 & \textbf{0.0279} & 24.55\%\\
& M@10 & 0.0097 & 0.0095 & 0.013 & 0.0139 & 0.02 & 0.0163 & \underline{0.0264} & 0.0003 & 0.0082 & 0.0131 & 0.0245 & 0.0112 & \textbf{0.0318} & 20.45\%\\
& M@20 & 0.0106 & 0.0102 & 0.0145 & 0.0158 & 0.0225 & 0.0198 & \underline{0.0296} & 0.0004 & 0.0095 & 0.0145 & 0.0271 & 0.0128 & \textbf{0.352} & 18.92\%\\

    \hline
    \multirow{3}{*}{ML-100K}

& M@5 & 0.0032 & 0.0036 & 0.0062 & 0.0063 & 0.01 & 0.0049 & 0.009 & 0.0017 & 0.0015 & 0.0039 & \underline{0.0186} & 0.0063 & \textbf{0.0209} & 13.37\%\\
& M@10 & 0.0041 & 0.004 & 0.0077 & 0.0079 & 0.0128 & 0.0071 & 0.0115 & 0.0022 & 0.0023 & 0.0053 & \underline{0.0225} & 0.0082 & \textbf{0.0267} & 18.67\%\\
& M@20 & 0.0048 & 0.0047 & 0.0091 & 0.0093 & 0.0154 & 0.0093 & 0.014 & 0.0027 & 0.0032 & 0.0065 & \underline{0.0259} & 0.0099 & \textbf{0.0305} & 17.76\%\\

    \bottomrule
  \end{tabular}
  \label{table:baseline2}
}
\vspace{-10pt}
\end{table*}

\subsection{Addition Comparison with Baselines}
\label{app:baselines}
We also show how TGODE compares with other baselines in terms of MRR@$k$ ($k$=5, 10, 20), and the results are shown in Table \ref{table:baseline2}.
As can be observed from the results, our model comprehensively outperforms the existing baseline on the MRR evaluation metric, demonstrating the superiority of our approach.

\subsection{Parameter Analysis}
\label{app:parameter}
To investigate the potential benefits of utilizing multiple propagation layers, we varied the number of layers in the GNN. Specifically, we conduct experiments on three different datasets, setting layer numbers in the range of $\{1,2,3,4,5\}$. 

\begin{figure}[t]
    \centering
    \includegraphics[width=\linewidth]{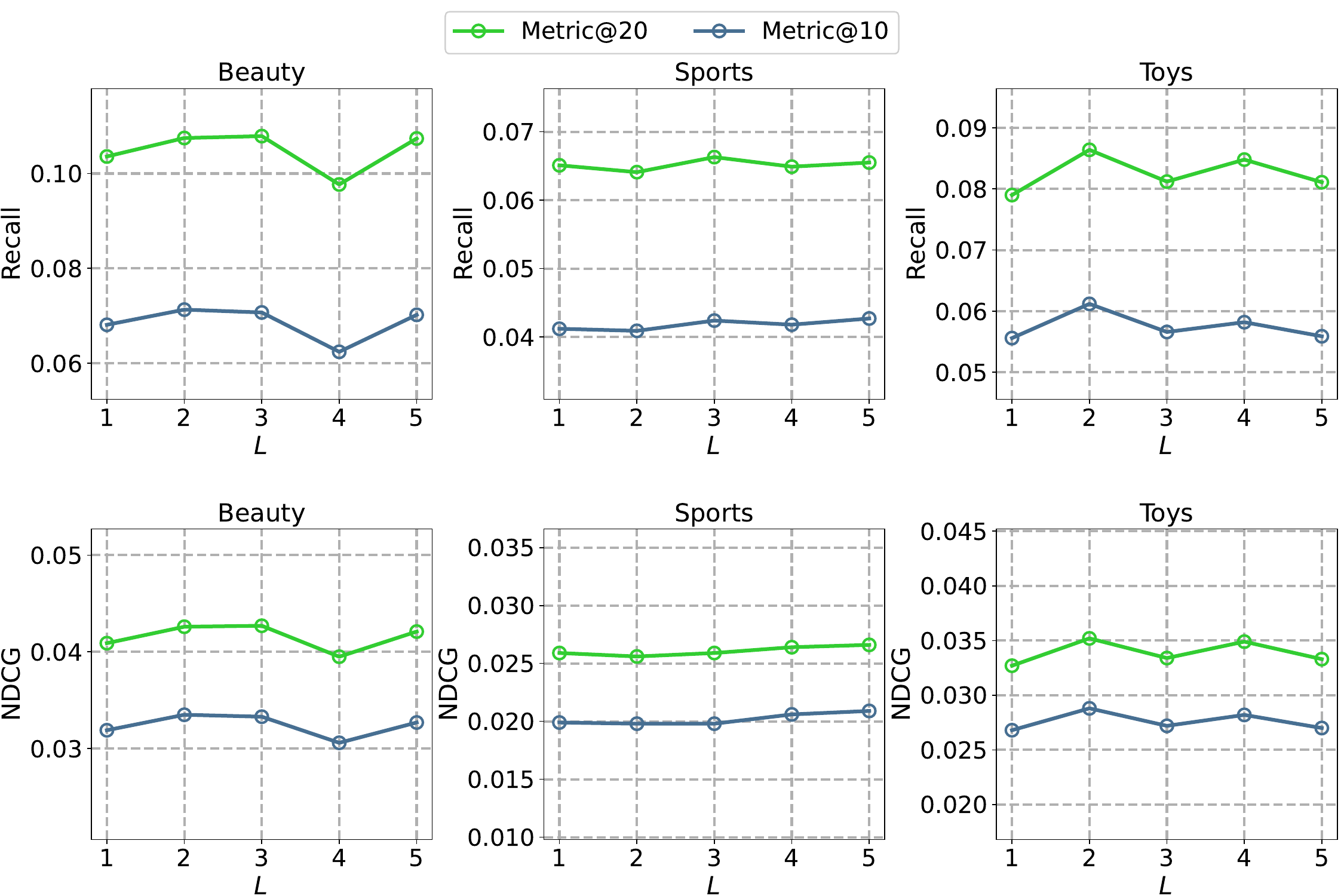}
\vspace{-20pt}
    \caption{The impact of hyperparameter $L$ on Beauty, Sports and Toys data.}
    
    \label{fig:hyper}
    \vspace{-8pt}
\end{figure}

The results are shown in the Figure \ref{fig:hyper}. TGODE achieves optimal performance when the number of layers is set to 2 or 3 on the Beauty and Sports datasets. This is because as the number of layers increases, the model's ability to handle complex data improves, which leads the model to capture richer information. However, as the layer continues to increase, the performance of the model deteriorates. This is due to the phenomenon of over-smoothing that occurs when using excessively deep layers, which degrades the model's performance. Experimental results confirm that setting the number of layers to 2 yields satisfactory performance across three datasets.
        
\vspace{-10pt}

\subsection{Related Work}

\subsubsection{Sequential Recommendation} The purpose of SR is to formulate and predict users' sequential interaction behaviors, capturing temporal dependencies and patterns. Current methodologies for SR can be primarily categorized into markov chains \cite{rendle2010factorizing, he2016fusing}, RNN-based \cite{hidasi2015session, yu2016dynamic, quadrana2017personalizing, li2018learning}, graph neural network (GNN)-based \cite{wu2019session, wang2020global, chang2021sequential, pan2020star}, and Transformer-based \cite{kang2018self, fan2021lighter, fan2022sequential, li2020time, hou2022core, sun2019bert4rec, chong2023ct4rec} approaches. Early works adopt markov chains to consider long-term dependencies and capture item-item sequence dependencies in SR. With the proliferation of deep learning techniques, the widespread adoption of RNN and GNN technologies has greatly facilitated the modeling of SR tasks by effectively excavating the information inherent in sequential and graph structures. For example, SRGNN \cite{wu2019session} utilizes graph attention networks to model individual sessions, enabling the representation of both the user's global preferences and current interests. The Transformer's multi-head attention mechanism and positional encoding enable it to directly and efficiently model global dependencies within sequences, thereby benefiting sequential recommendation tasks. MAERec \cite{ye2023graph} utilizes Transformer and graph-masked autoencoder to alleviate the problem of requiring high-quality graph embeddings in self-supervised learning. 

\subsubsection{Diffusion Models} The diffusion model has grown into a powerful deep learning model in recent years, and significant achievements have been made in related work in multiple aspects, either by optimizing model performance through practice \cite{song2020denoising} or by theoretically increasing model capacity \cite{lu2022maximum}. In applications, diffusion models dominate multiple challenging interdisciplinary tasks, including computer vision \cite{luo2021score,zhang2023adding}, natural language processing \cite{li2022diffusion}, multi-modal modeling \cite{huang2023collaborative}, and other interdisciplinary applications. For the field of recommender systems, approaches utilizing diffusion models \cite {liu2023diffusion, yang2024generate, wang2023diffusion} have also achieved superior performance with their excellent denoising and generation capabilities. For example,  DiffuASR \cite{liu2023diffusion} mitigates data sparsity and long-tail user issues in SR by employing a diffusion-based pseudo sequence generation framework and sequence U-Net model. DreamRec \cite{yang2024generate} utilizes guided diffusion to generate an ideal positive sample, eliminating negative samples and accurately capturing the user's real preferences. 

\subsubsection{Neural ODEs} Neural ODEs are powerful tools for handling dynamic system modeling and sequence modeling, with significant advantages in continuous modeling, flexible depth, and memory efficiency. Given its advantages in spatiotemporal sequences, ode has been widely applied in related aspects, e.g., time series prediction \cite{jin2022multivariate, gao2024egpde} and traffic flow forecasting \cite{zhong2023attention, fang2021spatial}. In the field of recommendation systems, ODE's time modeling ability is also well suited to user interaction sequences. For graph recommendation, GDERec \cite{qin2024learning} employs two customized GNNs trained alternately in an autoregressive manner to model and predict the evolution of user preferences, enabling tracking of the underlying system's dynamics under irregular observations. While for SR, considering the continuity of dynamic user preferences, GNG-ODE \cite{guo2022evolutionary} extends the idea of neural ODEs to continuous time session graphs and proposes a method to align the update time steps of time session graphs.

\end{document}